\documentclass[twocolumn,showpacs,preprintnumbers,amsmath,amssymb,unsortedaddress]{revtex4-2} 

\usepackage{graphicx,color}
\usepackage{dcolumn}
\usepackage{bm}
\usepackage{comment}
\usepackage{ulem} 
\usepackage{amsmath,amssymb}

\begin{document}


\title{
Theory of polymers in binary solvent solutions: mean-field free energy and phase behavior 
}
\author{Davide Marcato}
\email{dmarcato@sissa.it}
\affiliation{Scuola Internazionale Superiore di Studi Avanzati (SISSA), Via Bonomea 265, 34136 Trieste, Italy}
\author{Achille Giacometti}
\email{achille.giacometti@unive.it}
\affiliation{Dipartimento di Scienze Molecolari e Nanosistemi, Universit\`a Ca' Foscari Venezia, 30123 Venezia, Italy}
\affiliation{European Centre for Living Technology (ECLT) Ca' Bottacin, 3911 Dorsoduro Calle Crosera, 30123 Venezia, Italy}
\author{Amos Maritan}
\email{amos.maritan@unipd.it}
\affiliation{Laboratory of Interdisciplinary Physics, Department of Physics and Astronomy ``G. Galilei'', University of Padova, Padova, Italy and INFN, Sezione di Padova, via Marzolo 8, 35131 Padova, Italy}
\author{Angelo Rosa}
\email{anrosa@sissa.it}
\affiliation{Scuola Internazionale Superiore di Studi Avanzati (SISSA), Via Bonomea 265, 34136 Trieste, Italy}


\date{\today}

\begin{abstract}
We present a lattice model for polymer solutions, explicitly incorporating interactions with
a bath of solvent and cosolvent 
molecules.
By exploiting the well-known analogy between polymer systems and the $O(n)$-vector spin model in the limit $n \to 0$, we derive an exact field-theoretic expression for the partition function of the system.
The latter is then evaluated at the saddle point, providing a mean-field estimate of the free energy.
The resulting expression, which conforms to the Flory-Huggins type, is then used to analyze the system's stability with respect to phase separation, complemented by a numerical approach based on convex hull evaluation.
We demonstrate that this simple lattice model can effectively explain the behavior
of a variety of seemingly unrelated polymer systems, which have been predominantly investigated in the past only through numerical simulations.
This includes both, single-chain and multi-chain, solutions.
Our findings emphasize the fundamental, mutually competing, roles of solvent and cosolvent in polymer systems. 
\end{abstract}

\pacs{}


\maketitle

\section{Introduction}\label{sec:Intro} 

The coil-globule transition of a polymer chain in a bath of solvent molecules is one of the paradigmatic example of phase-transitions~\cite{KhokhlovGrosbergBook1994,Rubinstein2003,Doi2013soft}.
As the chemical affinity between the polymer and the solvent decreases, the solvent conditions are gradually moved from {\it good} to {\it poor} and the polymer tends to collapse from a randomly extended coil to a structureless globule, the analogue of a gas-liquid transition.
The same collapse can be also obtained at a fixed solvent quality and upon lowering the temperature~\cite{Stockmayer1960,Lifshitz1976,LifshitzRMP1978,TanakaNature1979,TanakaPRL1980,JanninkDesCloizeaux1990}.
In theoretical descriptions, however, the solvent microscopic degrees of freedom are usually traced out to obtain an effective polymer-polymer interaction, usually referred to as ``implicit solvent'' description.
While this approach is very convenient for analytical studies, it falls short in some cases where a detailed description of the solvent is required~\cite{Dongmo2024}.
Indeed, recent numerical studies~\cite{Huang2021,Garg2023} have shown that even a simple Lennard-Jones polymer chain in explicit solvent displays a non-trivial re-entrant collapse that defies the conventional interpretation based on the solvent quality.
Another striking example of this is the phenomenon known as {\it cosolvency}, where two different solvents compete with one another~\cite{Bharadwaj2021} thus triggering a complex phase behavior of the polymer chain~\cite{KremerNatCom2014,Freed2015,Sommer2018,Sommer2022cononsolvency,Zhang2024,Meng2024} that is still not fully understood~\cite{Bharadwaj2020}.
Understanding this mechanism in this, relatively simple, case would be of paramount importance for protein stability~\cite{Canchi2013,England2011}.
A counterpart of this phenomenon also occurs in the presence of two solvents, each of which separately is a good solvent for the polymer but their combination triggers the collapse of the polymer chain, and it is usually referred to as {\it co-nosolvency}.
Finally a different, albeit clearly related, situation occurs in the {\it polymer-assisted condensation} where phase separation of two miscible liquids is induced by a preferential attachment of a polymer chain to one of the solvents~\cite{SchiesselPAC2022}, a phenomenon of outmost importance in the nucleation of biomolecular condensates~\cite{Hyman2014,Berry2018,Pappu2023}.
When these situations occur in the presence of multiple chains, the behaviour is clearly even more complex as self-assembly can occur in the temperature-density plane~\cite{Olmsted1998,Arcangeli2024phase}.

Rather surprisingly, the three phenomenology described above have been discussed nearly always separately in the literature notwithstanding the evident analogies between them.
As a result, a general ``big picture'' is lacking, in spite of the significant numerical and theoretical work that has been profused in this framework~\cite{Polson2004,PolsonRiskJCP2009,Heyda2013,KremerNatCom2014,ZhaoKremer-Macromolecules2020,Huang2021,SchiesselPAC2022,Garg2023,Freed2015,Sommer2018, Sommer2022cononsolvency,Zhang2024, Meng2024}.

The aim of the present study is to address these issues within a unified theoretical framework.
Motivated by these considerations, and contrary to the conventional wisdom reckoning the above phenomenology as too complex and too specific to be investigated within a single analytical theory, in this paper we introduce a general $O(n)$-vector spin model for polymer solutions on the $d$-dimensional hypercubic lattice interacting with an explicit bath of two competing solvent species, generically referred to as {\it solvent} and {\it cosolvent}, and derive the exact partition function of the lattice model via a field theory.
Our approach builds on a classical scheme for a single polymer chain~\cite{Doniach1996}, recently generalized to multiple chains~\cite{Marcato2023}, but it extends it to incorporate the {\it explicit} presence of solvent and cosolvent.
We then provide a mean-field solution of this model that reproduces several phenomenological Flory-Huggins free energies~\cite{Flory1942,Huggins1942} in the appropriate limits. 
When combined with a graphical convex hull scheme, our mean-field theory is then shown to be able to recapitulate and rationalize the numerical results relative to the systems alluded early above, namely:
\begin{enumerate}
\item
The {\it re-entrant} collapse behavior for a single polymer chain immersed in a solvent triggered by strong monomer-solvent affinity~\cite{Huang2021,Garg2023}.
In the same context, we consider also the case of multi-chain systems.
\item
The polymer {\it co-nonsolvency} occurring in a ternary mixture with one polymer species and two solvents, where each solvent is good for the polymer but the disposed combination of the two turns out to be not~\cite{KremerNatCom2014,Freed2015,Sommer2018,Sommer2022cononsolvency,Zhang2024,Meng2024}.
\item
The {\it polymer-assisted condensation}~\cite{SchiesselPAC2022} where phase separation is observed in a two-component liquid mixture, induced by the presence of a polymer chain, as in nucleation of biomolecular condensates~\cite{Hyman2014,Berry2018,Pappu2023}.
\end{enumerate}
Our unified analysis suggests that these three phenomena, that have been studied as distinct entities in the past, are just particular cases of the same ``physics''.

The paper is structured as follows.
In Section~\ref{sec:ModelMethods}, we derive the $O(n)$-vector spin field theory (Sec.~\ref{sec:MicroscopicModel-Def}), that can be solved within a mean-field approximation (Sec.~\ref{sec:MicroscopicModel-MF}) leading to the free energy of the system (Sec.~\ref{sec:MicroscopicModel-FreeEn}).
In Sec.~\ref{sec:PhaseStability}, we derive the equations describing the stability of the free energy for single-chain systems with respect to phase separation (Sec.~\ref{sec:PhaseStability-SingleChain}) and the numerical convex hull construction (Sec.~\ref{sec:ConvexHull-GibbsTriangle}) for the efficient exploration of phase behavior as a function of the strength of the interactions between the different molecular species.
In Sec.~\ref{sec:Results}, we describe the main results following the application of our theory to the three distinct systems introduced above.
Finally, in Sec.~\ref{sec:DiscConcl} we provide a brief discussion of our formalism and of its applications considered here as well as of possible extensions of our work in the future.
The Supplemental Material (SM~\cite{SMnote}) adds some notes showing that classical, textbooks binary mixtures appear as particular cases of our formalism, together with a detailed account of how to derive the equations for phase stability in multi-chain systems as well as a few additional figures that complement the ones in the main text of the paper.

\section{Polymer-solvent lattice model}\label{sec:ModelMethods}

The first two Secs.~\ref{sec:MicroscopicModel-Def} and~\ref{sec:MicroscopicModel-MF} contain the necessary definitions of the main quantities describing the systems of interest, the formulation of the partition function and its field-theoretic representation.
All technical details can be skipped on first reading, and the reader can move to Sec.~\ref{sec:MicroscopicModel-FreeEn} where the (approximate) free energy of the system is presented and discussed in detail.

\subsection{Definition and exact field-theoretic formulation}\label{sec:MicroscopicModel-Def}

\begin{figure}
\centering
\includegraphics[scale=0.45]{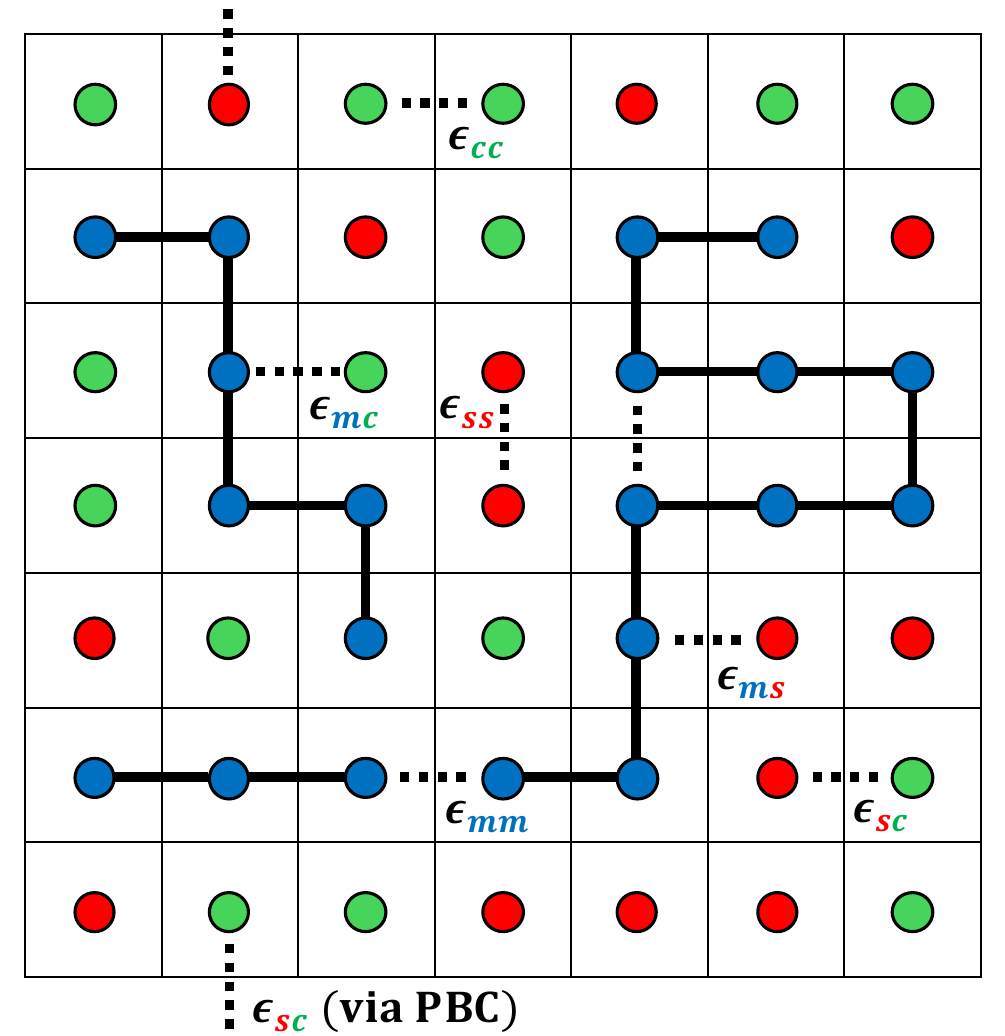}
\caption{
Illustration of a particular configuration on the square lattice ($d = 2$), with $N=7\times 7=49$ sites.
Monomers and the bonds joining them into polymer chains are represented, respectively, by blue dots and solid black lines.
Solvent and cosolvent molecules are represented by red and green dots, respectively. 
Some illustrative examples of nearest-neighbour interactions between molecular species (including one {\it via} periodic boundary conditions) are represented as dashed lines.
In the example here (Eq.~\eqref{eq:GranCanonicalZ}): $N_p = 3$, $N_b = 17$, $N_s = 14$.
}
\label{fig:ModelConformation}
\end{figure}

The model generalizes our field-theoretic formalism~\cite{Marcato2023} for self-interacting polymers on the $d$-dimensional hypercubic lattice by introducing explicit solvent
and cosolvent 
molecules.
By denoting with $N$ the total number of sites of the lattice and by taking the elementary lattice step as our unit of length, we assume, for simplicity, that monomers ($m$), solvent ($s$)
and cosolvent ($c$) 
molecules are of the same ``size'' equal to one single lattice site (see Fig.~\ref{fig:ModelConformation}).
In order to enforce the natural constraint of excluded volume, each lattice site is occupied by one single molecule (be it a monomer, a solvent
or a cosolvent 
one), while two
(non-bonded) 
molecules sitting at nearest-neighbor lattice positions interact via suitable pair interactions (so setting the stage for modeling the systems of Refs.~\cite{KremerNatCom2014,Huang2021,SchiesselPAC2022,Garg2023}, see below).

As seen in~\cite{Marcato2023} and for mathematical convenience, we adopt a {\it grand canonical} ensemble approach with given fugacities $\kappa_p$, $\kappa_b$ and $\kappa_s$ that fix the expectation values of the {\it total number} of, respectively, polymer chains $N_p$,
bonded monomer pairs 
$N_b$ and solvent molecules $N_s$.
As a result, the systems we consider are tipically {\it polydisperse}.
Moreover, with no empty sites the total number of cosolvent molecules in a given configuration is $N -  (N_b + N_p) - N_s$. 
Under these assumptions, with $\beta=1/(k_BT)$ being the inverse thermal energy at temperature $T$ ($k_B$ is the Boltzmann constant) the grand canonical partition function of the system reads
\begin{eqnarray}\label{eq:GranCanonicalZ}
Z = \sum_{\{\mathcal{C}\}} \kappa_p^{N_p} \kappa_b^{N_b} \kappa_s^{N_s}e^{ -\beta (\frac12 \sum_{i\neq j}\epsilon_{ij}N_{ij} + \sum_{i}\epsilon_{ii}N_{ii}) } \, ,
\end{eqnarray}
where the sum is for all possible configurations $\{ {\mathcal C} \}$,
and $N_{ij}$ denotes the total number of non-bonded pairs of species $i$ and $j$ sitting at nearest-neighbor lattice positions (including also neighbors {\it via} periodic boundary conditions, Fig.~\ref{fig:ModelConformation}) with corresponding pair interactions $\epsilon_{ij}$ (where $i,\,j \in \{ m, s, c \}$). 
Notice that
for simplicity (see however footnote~\cite{FlexibilityNote}), 
chains are modeled as fully flexible.

A first step towards the evaluation of $Z$ is done by noticing that the $N_{ij}$'s are not independent quantities, as it is easy to check that
\begin{eqnarray}
N_{mc} & = & 2d N_p + 2(d-1)N_b - 2N_{mm} - N_{ms} \, , \label{eq:DefineNmc} \\
N_{sc} & = & 2dN_s - N_{ms} - 2N_{ss} \, , \label{eq:DefineNsc} \\
N_{cc} & = & dN - 2dN_p -(2d-1)N_b-2dN_s \nonumber\\
& & + N_{mm} + N_{ss} + N_{ms} \, . \label{eq:DefineNcc}
\end{eqnarray}
Then, by Eqs.~\eqref{eq:DefineNmc}-\eqref{eq:DefineNcc} and by redefining fugacities and pair interactions as the following:
\begin{eqnarray}
\tilde{\kappa}_p & = & \kappa_p \, e^{-\beta(2d\epsilon_{mc}-2d\epsilon_{cc})} \, , \label{eq:RescaleKp}\\
\tilde{\kappa}_b & = & \kappa_b \, e^{-\beta(2(d-1)\epsilon_{mc} - (2d-1)\epsilon_{cc})} \, , \label{eq:RescaleKb}\\
\tilde{\kappa}_s & = & \kappa_s \, e^{-\beta(2d\epsilon_{sc} - 2d\epsilon_{cc})} \, , \label{eq:RescaleKs}\\
\tilde{\epsilon}_{mm} & = & \epsilon_{mm} - 2\epsilon_{mc} + \epsilon_{cc} \, , \label{eq:RescaleEmm}\\
\tilde{\epsilon}_{ss} & = & \epsilon_{ss} - 2\epsilon_{sc} + \epsilon_{cc} \, , \label{eq:RescaleEss}\\
\tilde{\epsilon}_{ms} & = & \epsilon_{ms} - \epsilon_{mc} - \epsilon_{sc} + \epsilon_{cc} \, , \label{eq:RescaleEms}
\end{eqnarray}
$Z$ can be rewritten as
\begin{eqnarray}\label{eq:RewriteZ}
Z & = & e^{-\beta d\epsilon_{cc}N}\sum_{\{\mathcal{C}\}} \tilde{\kappa}_p^{N_p} \tilde{\kappa}_b^{N_b} \tilde{\kappa}_s^{N_s}e^{ -\beta (\tilde{\epsilon}_{mm}N_{mm} + \tilde{\epsilon}_{ms}N_{ms} + \tilde{\epsilon}_{ss}N_{ss})} \nonumber\\
& = & e^{-\beta d\epsilon_{cc}N} Z_0 \, ,
\end{eqnarray}
where $Z_0$ is the grand canonical partition function of the equivalent system where only $mm$, $ss$ and $ms$ interactions appear while the various interactions involving the cosolvent are set $=0$.
Therefore, in the following, we concentrate on the evaluation of $Z_0$.

The grand canonical ensemble approach allows us to take advantage of the well known mapping between polymer systems and the $n \to 0$ limit of the $O(n)$-vector model for interacting spins~\cite{deGennes1972,deGennesBook,Orland1985,Bascle1992,Doniach1996,Marcato2023}.
In fact, by introducing~\cite{Marcato2023} at each lattice position $\mathbf x$ the $n$-component vector ${\mathbf S}(\mathbf x) \equiv (S^1(\mathbf x), S^2(\mathbf x), ..., S^n(\mathbf x))$ with the {\it internal product}
$
{\mathbf S}(\mathbf x) \cdot {\mathbf S}(\mathbf x^{\prime}) \equiv \sum_{i=1}^n S^i\!(\mathbf x) S^i\!(\mathbf x^{\prime})
$
between any two vectors associated to lattice points $\mathbf x$ and $\mathbf x^{\prime}$ and by defining the {\it trace} operation (denoted by the symbol $\langle ... \rangle_0$) through the formal rules:
\begin{eqnarray}
\langle 1 \rangle_0 & = & 1 \, , \label{eq:TraceProperties-a}\\
\langle S^i \rangle_0 & = & 0 \, , \label{eq:TraceProperties-b}\\
\langle S^i S^{j} \rangle_0 & = & \delta_{ij} \, , \label{eq:TraceProperties-c}\\
\langle S^{i_1} S^{i_2} \, ... \, S^{i_k} \rangle_0 & = & 0 \, , \, \, \, \text{if $k \geq 3$} \, , \label{eq:TraceProperties-d}
\end{eqnarray}
with $S$-vectors at different sites being independent of each other under the same trace operation, the following identity holds (compare to Eq.~(32) in~\cite{Marcato2023}) 
\begin{widetext}
\begin{eqnarray}\label{eq:PartitionFunct-sm}
Z_0
& = &
\int \prod_{\sigma}\mathcal{D}\psi_{\sigma} \, e^{ -\frac12 \sum_{\sigma}\sum_{{\vec x}, {\vec x}'} \Delta^{-1}({\mathbf x}, {\mathbf x}') \, \psi_\sigma({\mathbf x}) \, \psi_\sigma({\mathbf x}') } \nonumber\\
& & \times \lim_{n \to 0} \bigg\langle \prod_{\mathbf x} \bigg( 1 + H_p(\mathbf x)S^1\!({\mathbf x}) + H_s(\mathbf x) (S^1\!({\mathbf x}))^2 \bigg) \, e^{ \frac12 \sum_{\mathbf x, \mathbf x^{\prime}} \Delta(\mathbf x, \mathbf x^{\prime}) \, h(\mathbf x) h(\mathbf x^{\prime}) \, {\mathbf S}(\mathbf x) \cdot {\mathbf S}(\mathbf x^{\prime}) } \bigg\rangle_0 \, ,
\end{eqnarray}
\end{widetext}
where:
\begin{eqnarray}
\Delta({\mathbf x}, {\mathbf x}') & = & \left\{ \begin{array}{cl} 1 \, , & \mbox{if } |{\mathbf x} - {\mathbf x}'| = 1 \, \mbox{lattice step} \\ 0 \, , & \mbox{otherwise} \end{array} \right. , \label{eq:DefineDeltaxx'}\\
H_p(\mathbf x) & = & \sqrt{\tilde{\kappa}_p} \, e^{\frac{\sqrt{\beta(\tilde{\epsilon}_{ms}-\tilde{\epsilon}_{mm})}}2 \psi_{mm}(\mathbf x)+\frac{\sqrt{-\beta\tilde{\epsilon}_{ms}}}2 \psi_{ms}(\mathbf x)} \, , \label{eq:DefineHpx}\\
H_s(\mathbf x) & = & \tilde{\kappa}_s \, e^{\sqrt{\beta(\tilde{\epsilon}_{ms}-\tilde{\epsilon}_{ss})}\psi_{ss}(\mathbf x)+\sqrt{-\beta\tilde{\epsilon}_{ms}}\psi_{ms}(\mathbf x)} \, , \label{eq:DefineHsx}\\
h(\mathbf x) & = & \sqrt{\tilde{\kappa}_b} \, e^{\frac{\beta\tilde{\epsilon}_{mm}}{2}}e^{\frac{\sqrt{\beta(\tilde{\epsilon}_{ms}-\tilde{\epsilon}_{mm})}}{2}\psi_{mm}(\mathbf x)+\frac{\sqrt{-\beta\tilde{\epsilon}_{ms}}}{2}\psi_{ms}(\mathbf x)} \, , \nonumber\\
\label{eq:Definehx}
\end{eqnarray}
${\mathcal D}\psi_\sigma \equiv (2\pi)^{-N/2} \, (\det{\Delta})^{-1/2} \, \prod_{\vec{x}}d\psi_\sigma(\mathbf x)$ is the measure associated to the auxiliary scalar fields $\psi_\sigma = \psi_\sigma(\mathbf x)$ with $\sigma = \{ mm, ms, ss \}$ ({\it i.e.}, there are $3$ scalar fields per each lattice site) and -- importantly! -- the ``$\lim_{n \to 0}$'' operation is required~\cite{Marcato2023} to rule out all contributions to the partition function that include chain topologies different from the linear one.
Then, the last step consists in ``removing'' the dependence on the $\mathbf S$-vectors in the last term of Eq.~\eqref{eq:PartitionFunct-sm} in favor of the vector field ${\boldsymbol \varphi}(\mathbf x) \equiv (\varphi^1(\mathbf x), \varphi^2(\mathbf x), ..., \varphi^n(\mathbf x))$ with the associated measure
\begin{equation}\label{eq:DPhi-Def}
{\mathcal D}\varphi \equiv (2\pi)^{-nN/2} \left( \det{\Delta} \right)^{-n/2} \prod_{\mathbf x} d{\boldsymbol \varphi}(\mathbf x) \, ,
\end{equation}
by means of a standard Hubbard-Stratonovich transformation~\cite{HubbardPRL1959,Chaikin00}.
After some manipulations, and up to an unimportant prefactor, we finally get:
\begin{equation}\label{eq:fullz}
Z_0 = \lim_{n \to 0} \int \prod_{\mathbf x} \prod_{\sigma}d\psi_{\sigma}(\mathbf x) \int \prod_{\mathbf x} d{\boldsymbol \varphi}(\mathbf x) \, e^{ -A - B + \sum_{\mathbf x} \ln \left[ 1 + C\right] } \, ,
\end{equation}
where:
\begin{eqnarray}\label{eq:Apsi-Amu-Bphimu}
A
& = & \frac12 \sum_{\sigma}\sum_{\mathbf x, \mathbf x^{\prime}} \Delta^{-1} \! (\mathbf x, \mathbf x^{\prime}) \, \psi_{\sigma}(\mathbf x) \, \psi_{\sigma}(\mathbf x^{\prime}) \, , \nonumber\\
B
& = & \frac12 \sum_{\mathbf x, \mathbf x^{\prime}} \Delta^{-1} \! (\mathbf x, \mathbf x^{\prime}) \, {\boldsymbol \varphi}(\mathbf x) \cdot {\boldsymbol \varphi}(\mathbf x^{\prime}) \, , \nonumber\\
C
& = & H_s(\mathbf x) + \frac{h^{2}(\mathbf x)}{2} \,  |{\boldsymbol \varphi}(\mathbf x)|^2 + H_p(\mathbf x) \, h(\mathbf x) \, \varphi^1(\mathbf x)\, . \nonumber\\
\end{eqnarray}
The field-theoretic expression~\eqref{eq:fullz} cannot be evaluated directly.
It is, however, particularly amenable to a systematic expansion around the saddle-point~\cite{Orland1985,Bascle1992,Doniach1996,Marcato2023} that will be outlined in the next Sec.~\ref{sec:MicroscopicModel-MF}.

\subsection{Mean-field (saddle-point) formulation}\label{sec:MicroscopicModel-MF}
In order to compute the first term of the saddle-point expansion (equivalent to the mean-field approximation), we differentiate the exponential in Eq.~\eqref{eq:fullz} with respect to each $\varphi^i(\mathbf x)$ and $\psi_{\sigma}(\mathbf x)$ and set the obtained expressions equal to $0$.
Then, we simplify the problem further by looking only for those solutions that both
(i) satisfy translational invariance
and
(ii) break the $O(n)$ symmetry of the vector field~\cite{EasyOnLimit}, {\it i.e.}
$
{\boldsymbol \varphi}({\mathbf x}) = (\varphi, 0, \dots, 0)
$
and
$
\psi_{\sigma}({\mathbf x}) = \psi_{\sigma} \label{eq:SearchPsiSol}
$
for every $\mathbf x$.
In the end, that leads to:
\begin{eqnarray}
\frac{\varphi}{2d}
& = & \frac{h^{2}\varphi + H_{p}h}{1+ H_s + \frac{h^2}{2}\varphi^2+H_p h \varphi} \, , \label{eq:phisp} \\
\frac{\psi_{mm}}{2d}
& = & \frac{\sqrt{\beta(\tilde{\epsilon}_{ms}-\tilde{\epsilon}_{mm})} \bigg(\frac{h^2}{2}\varphi^2 + H_p h \varphi\bigg)}{1+ H_s + \frac{h^2}{2}\varphi^2+H_p h \varphi} \, , \label{eq:psispmm} \\
\frac{\psi_{ss}}{2d}
& = & \frac{\sqrt{\beta(\tilde{\epsilon}_{ms}-\tilde{\epsilon}_{ss})} H_s}{1+ H_s + \frac{h^2}{2}\varphi^2+H_p h \varphi} \, , \label{eq:psispss} \\
\frac{\psi_{ms}}{2d}
& = & \frac{\sqrt{-\beta\tilde{\epsilon}_{ms}} \bigg(H_s + \frac{h^2}{2}\varphi^2 + H_p h \varphi\bigg)}{1+ H_s + \frac{h^2}{2}\varphi^2+H_p h \varphi} \, , \label{eq:psispms}
\end{eqnarray}
where $H_p$, $H_s$ and $h$ are the same quantities defined in Eqs.~\eqref{eq:DefineHpx}-\eqref{eq:Definehx}, computed in correspondence of the saddle-point.
In terms of the solutions~\cite{OnPhiPsiNotation}
$\varphi = \varphi(\tilde{\kappa}_p, \tilde{\kappa}_b, \tilde{\kappa}_s, \tilde{\epsilon}_{mm}, \tilde{\epsilon}_{ss}, \tilde{\epsilon}_{ms})$
and
$\psi_{\sigma} = \psi_{\sigma}(\tilde{\kappa}_p, \tilde{\kappa}_b, \tilde{\kappa}_s, \tilde{\epsilon}_{mm}, \tilde{\epsilon}_{ss}, \tilde{\epsilon}_{ms})$
of the mean-field Eqs.~\eqref{eq:phisp}-\eqref{eq:psispms}, the corresponding grand potential per lattice site, $\beta\Omega \equiv -\ln Z/N = \beta d \epsilon_{cc} -\ln(Z_0) / N$ (see Eq.~\eqref{eq:RewriteZ}), reads (up to an unimportant additive constant) as the following: 
\begin{widetext}
\begin{equation}\label{eq:completefreeene-sm}
\beta \Omega(\tilde{\kappa}_p, \tilde{\kappa}_b, \tilde{\kappa}_s, \tilde{\epsilon}_{mm}, \tilde{\epsilon}_{ss}, \tilde{\epsilon}_{ms}) = 
\beta d \epsilon_{cc} + \frac{\psi^2_{mm}}{4d} + \frac{\psi^2_{ms}}{4d} + \frac{\psi^2_{ss}}{4d} + \frac{\varphi^2}{4d} - \ln \bigg[ 1 +  H_s + H_p h \varphi + \frac{h^2}{2}\varphi^2 \bigg] \, .
\end{equation}
\end{widetext}
Eq.~\eqref{eq:completefreeene-sm}, alongside Eqs.~\eqref{eq:DefineHpx}-\eqref{eq:Definehx} and Eqs.~\eqref{eq:phisp}-\eqref{eq:psispms} calculated at the saddle-point, defines completely the thermodynamics of the system.
In particular, it is easy to derive the expressions,
\begin{eqnarray}
\phi_b & \equiv & \frac{\langle N_b \rangle}N = -\beta\kappa_b \frac{\partial \Omega}{\partial \kappa_b} = \frac{\varphi^2}{4d}\, , \label{eq:<phib>} \\
\phi_p & \equiv & \frac{\langle N_p \rangle}N = -\beta\kappa_p \, \frac{\partial \Omega}{\partial\kappa_p} = \frac{H_p h \varphi}{2\left(1+H_s + H_p h \varphi + \frac{h^2}{2}\varphi^2\right)} \, , \nonumber\\
\label{eq:<phic>} \\
\phi_m & \equiv & \phi_p + \phi_b = \frac{\frac{h^2}2 \varphi^2 + H_p h \varphi}{1+H_s + H_p h \varphi + \frac{h^2}{2}\varphi^2} \, , \label{eq:<phim>} \\
\phi_s & \equiv & \frac{\langle N_s \rangle}N = -\beta\kappa_s \frac{\partial \Omega}{\partial \kappa_s} = \frac{H_s}{1+H_s + H_p h \varphi + \frac{h^2}{2}\varphi^2} \, , \nonumber\\
\label{eq:<phis>}
\end{eqnarray}
for, respectively, the (mean) bond, chain, monomer and solvent fraction (or, density).

\subsection{Free energy of the system}\label{sec:MicroscopicModel-FreeEn}
Instead of dealing with the grand potential $\beta\Omega$~\eqref{eq:completefreeene-sm}, a more transparent characterization of the thermodynamics of the system is obtained in terms of the {\it free energy} per lattice site (in units of $k_BT = \beta^{-1}$),
\begin{equation}\label{eq:FreeEneExplicitSolvent-sm}
\beta f = \beta f(\phi_p, \phi_b, \phi_s) \equiv \beta\Omega + \phi_p \ln\kappa_p + \phi_b \ln\kappa_b + \phi_s\ln\kappa_s \, ,
\end{equation}
equivalent to the Legendre transform~\cite{Marcato2023} of $\beta\Omega$.
By expressing the fugacities~\eqref{eq:RescaleKp}-\eqref{eq:RescaleKs}
\begin{eqnarray}
\tilde{\kappa}_p & = & \frac{2\phi^2_p \, \,  e^{2\beta d(\tilde{\epsilon}_{mm}\phi_m + \tilde{\epsilon}_{ms}\phi_s)}}{(\phi_b-\phi_p)(1 - \phi_m - \phi_s)} \, , \\
\tilde{\kappa}_b & = & \frac{(\phi_b-\phi_p) \, e^{2\beta d(\tilde{\epsilon}_{mm}\phi_m + \tilde{\epsilon}_{ms}\phi_s) - \beta\tilde{\epsilon}_{mm}}}{2d\phi_b(1 - \phi_m - \phi_s)} \, , \\
\tilde{\kappa}_s & = & \frac{\phi_s \, e^{2\beta d(\tilde{\epsilon}_{ms}\phi_m + \tilde{\epsilon}_{ss}\phi_s)}}{1 - \phi_m - \phi_s} \, ,
\end{eqnarray}
as a function of densities $\phi_p$, $\phi_m$ and $\phi_s$ (Eqs.~\eqref{eq:<phic>}-\eqref{eq:<phis>}) and after a few simple manipulations, one gets the final expression,
\begin{widetext}
\begin{eqnarray}\label{eq:FreeEneExplicitSolvent}
\beta f(\phi_m, \phi_s, \ell)
& = & \beta d(\epsilon_{mm} - 2\epsilon_{mc} + \epsilon_{cc}) \, \phi_m^2 + \beta d (\epsilon_{ss} - 2\epsilon_{sc} + \epsilon_{cc})\, \phi_s^2 + 2\beta d(\epsilon_{ms}-\epsilon_{mc} - \epsilon_{sc} + \epsilon_{cc})\phi_m\phi_s \nonumber\\
& & + (1 - \phi_m - \phi_s) \ln(1 - \phi_m - \phi_s) + \frac{\phi_m}{\ell} \ln(\phi_m) + \phi_s \ln(\phi_s) \nonumber\\
& & + \phi_m \left[ \ln\!\left( \frac{(1-2/\ell)^{1-2/\ell} \, (2/{\ell}^2)^{1/\ell}}{ ((1-1/\ell)2d/e)^{1-1/\ell}} \right) - \beta\epsilon_{mm} \left(1-\frac1{\ell}\right) + 2d\beta\left(\epsilon_{mc}-\epsilon_{cc}\right) \right] \, \nonumber\\
& & + 2d\beta (\epsilon_{sc} - \epsilon_{cc})\phi_s + \beta d\epsilon_{cc}
\end{eqnarray}
\end{widetext}
as a function of monomer density ($\phi_m$), solvent density ($\phi_s$) and {\it mean} chain contour length ($\ell \equiv \phi_m / \phi_p$). 
Notice, in particular, that the Legendre transform allows us to switch to the ensemble where the populations of different species are fixed, yet polydispersity is still unavoidable.

Interestingly, $\beta f$~\eqref{eq:FreeEneExplicitSolvent} is of the familiar Flory-Huggins form~\cite{Flory1942,Huggins1942}. 
However, unlike most conventional presentations (for instance, see~\cite{Rubinstein2003}), the derivation of~\eqref{eq:FreeEneExplicitSolvent} presents two main advantages:
(i) it proceeds from a genuinely microscopic model;
(ii) it can be improved beyond the saddle-point through the systematic inclusion of higher-order corrections~\cite{Bawendi1988}.

It may be noticed that the mean-field expression Eq.~\eqref{eq:FreeEneExplicitSolvent} contains additional, non-trivial, contributions with respect to the usual Flory-Huggins formula, which are either linear in $\phi_m$, linear in $\phi_s$ or constant.
Among these terms it is easy to distinguish between energetic and entropic ones.
The energy terms are proportional to $\epsilon_{ij}$ (for some $i,j$) and they essentially correct for the overcounting of {\it mm} and {\it cc} interactions (the former by considering that only non-bonded monomers interact).
Perhaps, less obvious is the entropic term.
The latter can be further broken down into the sum of two terms: the first is $\phi_m(1-1/\ell)\ln(2d/e)$, which resembles the usual mean-field configurational entropy for a single chain on a lattice~\cite{Orland1985,Doniach1996}, while the other arises from the entropy related to polydispersity.
It is worth noticing the consequence of the linear (or constant) dependence of these terms upon the densities.
In fact, the free energy {\it variation} upon mixing (which is actually the central quantity of the usual Flory-Huggins theory) is defined~\cite{deGennesBook} as
$\Delta f(\phi_m,\phi_s,\ell) \equiv f(\phi_m,\phi_s,\ell) - [ \phi_m f(1,0,\ell) + \phi_s f(0,1,\ell) + (1-\phi_m-\phi_s)f(0,0,\ell) ]$: it is then evident that by this definition all the terms which are linear or constant do not contribute to $\Delta f$.
Therefore, for what concerns the free energy variation, our field-theoretic approach reproduces {\it exactly} the Flory-Huggins formula. 
Finally, the free energy $\beta f$~\eqref{eq:FreeEneExplicitSolvent} describes the thermodynamics of a {\it ternary} mixture~\cite{Tompa1949,Altena1982} of polymers, solvent and
cosolvent, 
with the constraint $\phi_m + \phi_s + \phi_c = 1$ where $\phi_c$ is the
cosolvent 
density; it reduces to known cases of {\it binary} mixtures when one species is absent (see the discussion in Sec.~\ref{sec:BinaryMixtures} in SM~\cite{SMnote}).
Importantly, the ratio $\phi_m/\ell$ tunes the chain number density: at fixed $\phi_m$ and in the thermodynamic limit $N \to \infty$, a finite $\ell$ corresponds to a (polydisperse) multi-chain solution, while $\ell \to \infty$ gives the single-chain limit~\cite{Binder2007}.
Eq.~\eqref{eq:FreeEneExplicitSolvent} is therefore valid for both, single and multiple chains, as it will be discussed further below.

\section{Phase stability and behavior: methods}\label{sec:PhaseStability}

\subsection{Single-chain systems}\label{sec:PhaseStability-SingleChain}
For a generic, complex ternary mixture of a single polymer chain ($\ell\to\infty$ in Eq.~\eqref{eq:FreeEneExplicitSolvent}), solvent and
cosolvent, 
we detail first the thermodynamics of separation in two phases (termed $I$ and $II$) in Sec.~\ref{sec:PhaseStability-SingleChain-2} while three-phase coexistence (termed $I$, $II$ and $III$) is outlined in Sec.~\ref{sec:PhaseStability-SingleChain-3}.
The generalization to multi-chain systems ($\ell<\infty$ in Eq.~\eqref{eq:FreeEneExplicitSolvent}) is then straightforward and we leave it to Sec.~\ref{sec:PhaseStability-MultiChains} in SM~\cite{SMnote}.

In order to understand the coming formalism, it ought to be stressed that the polymer coil state is equivalent to a stable mixed phase whereas the globule state is a phase-separate system where one phase has $\phi_m>0$ and the other(s) have $\phi_m=0$~\cite{Binder2007,deGennesBook}.
To clarify this point, consider for simplicity a binary polymer/
solvent 
mixture (Eq.~\eqref{eq:FreeEneExplicitSolvent} with $\phi_m + \phi_s = 1$): the critical point is for $\phi_m^\ast = 1/(1 + \sqrt{\ell})$~\cite{Rubinstein2003,Qian2022}, with the two branches of the binodal line lying to the left and to the right of it.
As $\ell \to \infty$, both $\phi_m^\ast$ and the binodal left branch $\to 0$, therefore the value of $\phi_m$ for the polymer-poor phase must also be $=0$ (see Fig.~1 in~\cite{Binder2007}).

\subsubsection{Two-phase coexistence}\label{sec:PhaseStability-SingleChain-2}
Phase $I$ is characterized by $\phi_m^I = 0$, $\phi_s^I>0$ and occupies volume $V^I$, while phase $II$ is characterized by $\phi_m^{II}>0$, $\phi_s^{II}>0$ and occupies volume $V^{II}$.
By generalizing standard arguments for binary mixtures~\cite{Qian2022}, these phases and their regions of stability are determined by minimizing the total free energy
\begin{equation}\label{eq:FreeEnePhaseSeparated}
V^I f(0, \phi^{I}_{s}) + V^{II}f(\phi^{II}_m, \phi^{II}_s) \, ,
\end{equation}
of the phase-separate system~\cite{NeglectSurfaceTension}, with additional constraints on volume and particle number of each species:
\begin{eqnarray}
V^I + V^{II} & = & V \, , \label{eq:2Phases-Constraints-a} \\
V^{II} \phi^{II}_m & = & V\phi_m \, , \label{eq:2Phases-Constraints-b} \\
V^I \phi^{I}_s + V^{II} \phi^{II}_s & = & V \phi_s \, , \label{eq:2Phases-Constraints-c}
\end{eqnarray}
where $V$ is the total volume of the system and $\phi_m$ and $\phi_s$ are the densities at which the system is prepared.
Then, by introducing the expressions
for the osmotic pressure
\begin{equation}\label{eq:DefineOsmoticPressure}
\Pi(\phi_m, \phi_s) \equiv \phi_m \frac{\partial f}{\partial \phi_m} + \phi_s \frac{\partial f}{\partial \phi_s} - f(\phi_m, \phi_s) \, ,
\end{equation}
and the chemical potential of the solvent
\begin{equation}\label{eq:DefineChemicalPot}
\mu_s(\phi_m, \phi_s) \equiv \frac{\partial f}{\partial \phi_s} \, ,
\end{equation}
%
minimization of Eq.~\eqref{eq:FreeEnePhaseSeparated} with constraints~\eqref{eq:2Phases-Constraints-a}-\eqref{eq:2Phases-Constraints-c} {\it via} standard Lagrange multipliers leads to 5 coupled equations with 3 constraints that, by some manipulations, give: 
\begin{eqnarray}
\Pi(0, \phi_s^I) & = & \Pi(\phi_m^{II}, \phi_s^{II}) \, , \label{eq:CoexistenceEqs1} \\
\mu_s(0, \phi_s^I) & = & \mu_s(\phi^{II}_m, \phi_s^{II}) \, , \label{eq:CoexistenceEqs2} \\
\phi_m^{II}(\phi_s^I - \phi_s) & = & \phi_m(\phi_s^I - \phi_s^{II}) \, . \label{eq:CoexistenceEqs3}
\end{eqnarray}
Eqs.~\eqref{eq:CoexistenceEqs1} and~\eqref{eq:CoexistenceEqs2} can be easily identified as the balance of the osmotic pressures and the chemical potentials of the solvent in the two phases~\cite{NoPolymerChemPotentialNote}, while Eq.~\eqref{eq:CoexistenceEqs3} is a generalization of the usual Maxwell lever rule for binary mixtures~\cite{Rubinstein2003}.
If, for a given pair $(\phi_m, \phi_s)$, a solution to Eqs.~\eqref{eq:CoexistenceEqs1}-\eqref{eq:CoexistenceEqs3} exists, the system minimizes its free energy by phase separating into two coexisting phases and is in a {\it biphasic} region (for the {\it triphasic} case, see Sec.~\ref{sec:PhaseStability-SingleChain-3} below); in polymer language, the chain collapses to a globule.
The solution to Eqs.~\eqref{eq:CoexistenceEqs1}-\eqref{eq:CoexistenceEqs3} is then used (Eqs.~\eqref{eq:2Phases-Constraints-a}-\eqref{eq:2Phases-Constraints-c}) to infer $V^I$ and $V^{II}$, so characterizing the two phases completely.

\subsubsection{Three-phase coexistence}\label{sec:PhaseStability-SingleChain-3}
In order to derive the conditions for the coexistence of three phases in single-chain systems we proceed similarly as for the case of two phases.
The starting point is again the total free energy of the phase-separated system (compare to Eq.~\eqref{eq:FreeEnePhaseSeparated}),
\begin{equation}\label{eq:3phasesFE}
V^I f(0, \phi_s^I) + V^{II} f(0, \phi_s^{II}) + V^{III} f(\phi_m^{III}, \phi_s^{III}) \, ,
\end{equation}
with the constraints (compare to Eqs.~\eqref{eq:2Phases-Constraints-a}-\eqref{eq:2Phases-Constraints-c}):
\begin{eqnarray}
V^I + V^{II} + V^{III} & = & V \, , \label{eq:3Phases-Constraints-a} \\
V^{III} \phi^{III}_m & = & V\phi_m \, , \label{eq:3Phases-Constraints-b} \\
V^I \phi_s^I + V^{II} \phi_s^{II} + V^{III} \phi_s^{III} & = & V \phi_s \, . \label{eq:3Phases-Constraints-c}
\end{eqnarray}
This time, the minimization procedures leads to a system of 7 equations (with 3 constraints), that can be rearranged into the following relations for the equilibrium densities:
\begin{eqnarray}
\Pi(0, \phi^{I}_s) & = & \Pi(0, \phi^{II}_s) \, , \label{eq:CoexistenceEqsTriPhase-a} \\
\Pi(0, \phi^{I}_s) & = & \Pi(\phi^{III}_m, \phi^{III}_s) \, , \label{eq:CoexistenceEqsTriPhase-b} \\
\mu_s(0, \phi_s^I) & = & \mu_s(0, \phi_s^{II}) \label{eq:CoexistenceEqsTriPhase-c} \\
\mu_s(0, \phi_s^I) & = & \mu_s(\phi^{III}_m, \phi_s^{III}) \, . \label{eq:CoexistenceEqsTriPhase-d}
\end{eqnarray}
Again, if a non-trivial solution to Eqs.~\eqref{eq:CoexistenceEqsTriPhase-a}-\eqref{eq:CoexistenceEqsTriPhase-d} exists, then the system minimizes its free energy by separating into three coexisting phases.
An interesting difference with respect to the biphasic case (compare to Eqs.~\eqref{eq:CoexistenceEqs1}-\eqref{eq:CoexistenceEqs3}) is that the solution to Eqs.~\eqref{eq:CoexistenceEqsTriPhase-a}-\eqref{eq:CoexistenceEqsTriPhase-d} {\it does not} depend explicitly on the {\it average densities} $\phi_m$ and $\phi_s$.

\begin{figure}
\includegraphics[width=0.45\textwidth]{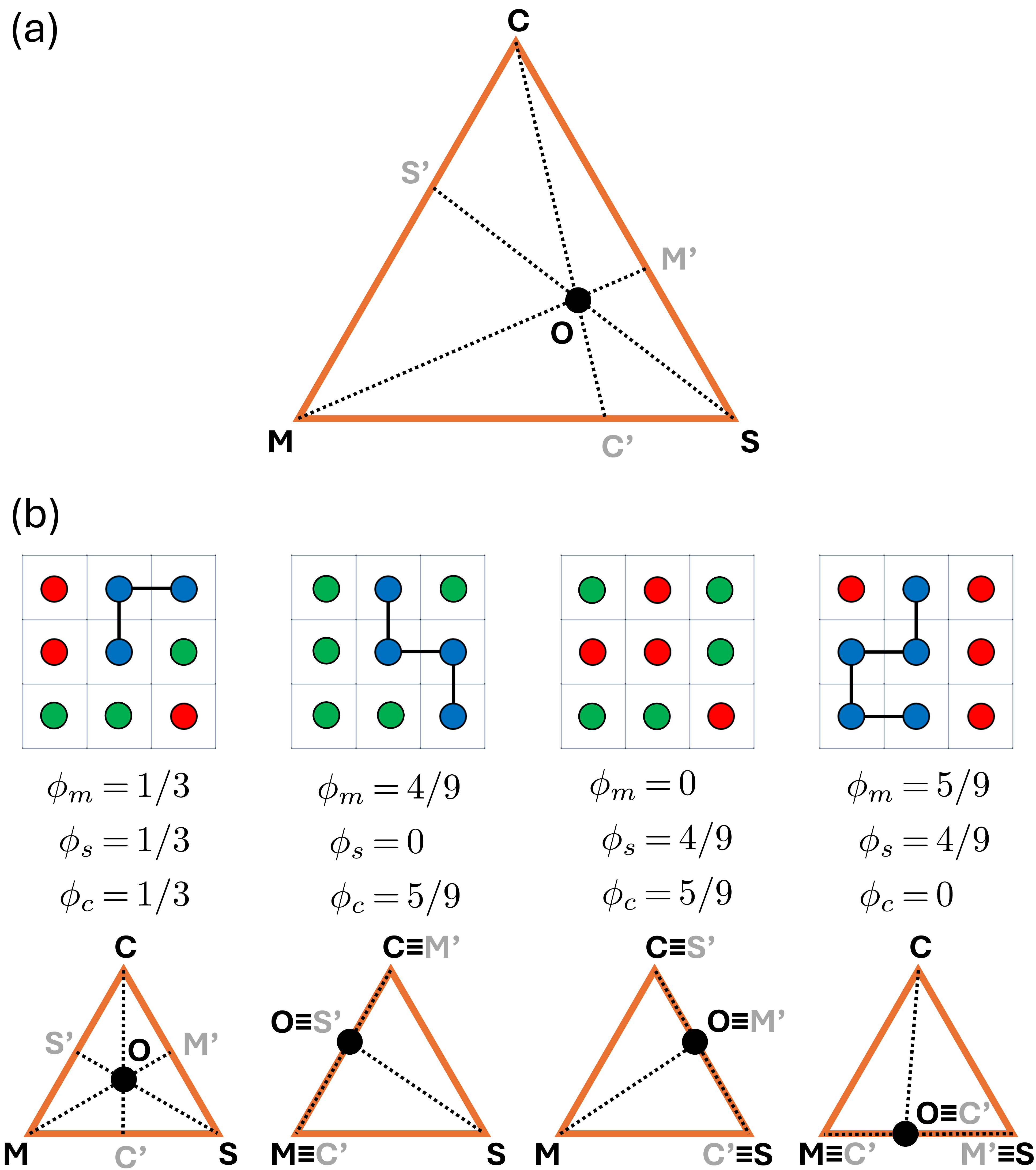}
\caption{
(a)
Gibbs triangle for a ternary mixture of monomers (vertex $M$), solvent molecules (vertex $S$) and 
cosolvent molecules (vertex $C$). 
A generic point $O$ inside the triangle stands for system's composition with densities: $\phi_m = |OM'|/|MM'|$, $\phi_s = |OS'|/|SS'|$ and $\phi_c = |OC'|/|CC'|$, where ``$|\cdot|$'' denotes the length of the segment.
By elementary geometry, it is easily seen that $\phi_m + \phi_s + \phi_c = 1$.
(b)
Illustrative cases (dots' color code is as in Fig.~\ref{fig:ModelConformation}), featuring (from left to right): equipopulation of the three species ($O$ is the barycenter) and when one species is absent.
}
\label{fig:GibbsTriangle}
\end{figure}

\subsection{Convex hull construction and Gibbs triangle}\label{sec:ConvexHull-GibbsTriangle}
Solutions to Eqs.~\eqref{eq:CoexistenceEqs1}-\eqref{eq:CoexistenceEqs3} for two-phase coexistence and Eqs.~\eqref{eq:CoexistenceEqsTriPhase-a}-\eqref{eq:CoexistenceEqsTriPhase-d} for three-phase coexistence require a non-trivial numerical procedure~\cite{Tompa1949,Altena1982,CodeLinkToGitHub}.
The task, however, can be greatly simplified by looking at the ``geometrical'' meaning~\cite{Wolff2011,Mao2019} of these equations, {\it i.e.} the evaluation of the {\it lower convex hull} (l.c.h.)~\cite{Preparata2012,Quickhull1996} of the free energy surface.
In particular, the regions where the l.c.h. differs from the free energy surface are those where phase separation occurs (again, analogously to the Maxwell construction in binary mixtures).
In order to identify those regions we follow~\cite{Wolff2011,Mao2019} and reconstruct the shape of the l.c.h. by the accurate triangulation procedure introduced therein.
We implement this method {\it via} the publicly available {\it Quickhull} package~\cite{Quickhull1996}, and get an estimated phase diagram of the system that is then used in concert with the numerical solutions of Eqs.~\eqref{eq:CoexistenceEqs1}-\eqref{eq:CoexistenceEqs3} and Eqs.~\eqref{eq:CoexistenceEqsTriPhase-a}-\eqref{eq:CoexistenceEqsTriPhase-d}.
The obtained stable-phase solutions are represented in terms of the characteristic {\it barycentric coordinates}~\cite{Tompa1949, Altena1982,Howarth1996} of the Gibbs triangle (Fig.~\ref{fig:GibbsTriangle}).
The described procedure is applied also to the identification of two- and three-phase coexistence in multi-chain systems (see Sec.~\ref{sec:PhaseStability-MultiChains} in SM~\cite{SMnote}).

\section{Results}\label{sec:Results}

\subsection{
Re-entrant collapse of polymer chains in solvent of varying quality 
}\label{sec:Results-SingleChain}
We start by applying our theory to the following problem~\cite{Huang2021}: a single chain described by the Kremer-Grest bead-spring polymer model~\cite{Kremer1990} in explicit solvent conditions, where $mm$ and $ss$ interactions are described by the same attractive Lennard-Jones interaction.
As the strength of the $ms$ attraction (also of the Lennard-Jones type) increases, the chain swells as expected in standard good solvent conditions.
However, by increasing the $ms$ attraction even further, the polymer is observed to fold back thus giving a re-entrant collapse (Fig.~\ref{fig:MDmodels}(a) in SM~\cite{SMnote}).

\begin{figure}
\includegraphics[width=0.46\textwidth]{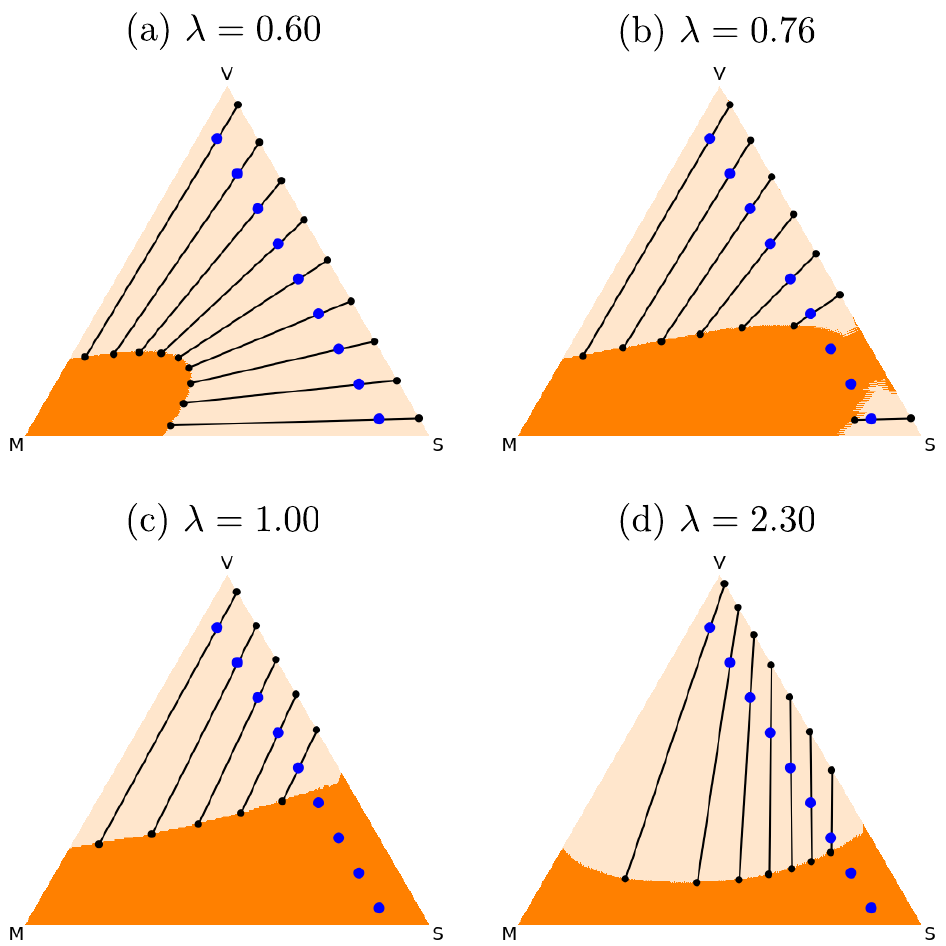}
\caption{
Phase behavior of a single chain in a one-solvent bath of varying quality (Eq.~\eqref{eq:FreeEneExplicitSolvent} for $d=3$, $\epsilon_{mm}=\epsilon_{ss}=-\epsilon<0$, $\epsilon_{ms}=-\lambda\epsilon$ ($\lambda>0$), $\epsilon_{cc}=\epsilon_{mc}=\epsilon_{sc}=0$ and temperature $k_BT/\epsilon=2.5$). 
The dark- and light-shaded areas correspond, respectively, to the stable (coil) and the biphasic (globule) region as identified from the convex hull procedure (see text for details).
Large blue dots correspond to $9$ chosen mean compositions of the system with the same $\phi_m=0.1$, while tiny black dots (connected by black lines) denote the compositions of the two stable phases in which the system separates.
The positions of the black dots are calculated by solving numerically Eqs.~\eqref{eq:CoexistenceEqs1}-\eqref{eq:CoexistenceEqs3}.
}
\label{fig:MeanFieldResultsT2.5}
\end{figure}

In order to rationalize this polymer behavior, consider Eq.~\eqref{eq:FreeEneExplicitSolvent} with $\ell \to \infty$, $\epsilon_{mm} = \epsilon_{ss} = -\epsilon<0$ and with $\epsilon_{ms}=-\lambda\epsilon<0$ where $\lambda$ tunes the $ms$ attraction.
In addition, as no other species was considered in the original set-up~\cite{Huang2021}, we fix $\epsilon_{mc} = \epsilon_{sc} = \epsilon_{cc} = 0$; noticeably, this allows us to ``interpret'' the cosolvent molecules as {\it lattice vacancies}.
To emphasize this aspect, in this as well as in Sec.~\ref{sec:PAC} on polymer-assisted condensation, the letter ``c'' for ``cosolvent'' is replaced by the letter ``v'' for ``vacancy'' in all quantities and figures of pertinence. 
Then, we focus on the phase stability of $\beta f$~\eqref{eq:FreeEneExplicitSolvent} as a function of densities $\phi_m$ and $\phi_s$, parameter $\lambda$ and temperature $T$ and, hereafter, in spatial dimensions $d=3$.
When the chosen densities (defining the mean composition of the system) do not belong to the stable region (as identified by the convex hull procedure), we solve numerically Eqs.~\eqref{eq:CoexistenceEqs1}-\eqref{eq:CoexistenceEqs3} and Eqs.~\eqref{eq:CoexistenceEqsTriPhase-a}-\eqref{eq:CoexistenceEqsTriPhase-d} to determine the stable phases and represent the corresponding phase separation by a solid black line (a.k.a. a {\it tie-line}) in the Gibbs triangle.
To fix the ideas, we consider the stability of $9$ representative coordinates $(\phi_m, \phi_s, \phi_v=1-\phi_m-\phi_s)$ in the Gibbs triangle with the same $\phi_m = 0.1$, $\phi_s \in [0.05-0.85]$ (large blue dots in Fig.~\ref{fig:MeanFieldResultsT2.5}).

We start by fixing the temperature $k_BT / \epsilon = 2.5$ and increase $\lambda$ systematically (Fig.~\ref{fig:MeanFieldResultsT2.5}): this corresponds to changing the solvent quality from ``poor" to ``good".
For $\lambda = 0.6$ (panel (a)), none of the points is stable as they all lay in the biphasic region (light shaded area).
This means that at this monomer concentration and temperature the system always phase separates in two different phases represented by the end small dots of each black line.
One of these phases (lying on the $SV$ edge) contains only solvent molecules and vacancies, mixed.
The other phase (lying on the edge of the dark shaded region) is a polymer globule mixed with solvent molecules and vacancies.
As $\lambda$ increases, the dark shaded region first touches the $SV$ edge, and two of the original state points become stable (panel (b) for $\lambda=0.76$) with the chain in the coil conformation as described earlier~\cite{Binder2007}.
Upon further increase of $\lambda$ more and more points are incorporated in the stable region up to $\lambda = 1.0$ (panel (c)); notice that for low $\phi_s$ the system still phase-separates.
Increasing $\lambda$ further (panel (d)), a new feature emerges: the extension of the stable region decreases and some of the former stable points switch back to phase separation.
This indeed represents the re-entrant globule phase described in~\cite{Huang2021} and it can be regarded as the counterpart of a similar phenomenon observed in colloids~\cite{Fantoni2015}.
Intriguingly, this re-entrance can be observed only for ``intermediate'' $\phi_s$.
At high solvent densities the system remains in a coil state (at least, up to the value $\lambda = 2.3$ considered here), while at low $\phi_s$ the polymer never experiences the coil-globule transition and the system remains in the biphasic region at all $\lambda$'s; the only significant modification is the compactness of the globule with the latter becoming more and more swollen as $\lambda$ increases.
Once more, this matches the findings of computer simulations by Garg {\it et al.}~\cite{Garg2023}, that were performed at a much lower solvent density than Ref.~\cite{Huang2021}.
There, however, the authors claimed a direct ``transition'' from a compact globule to a less compact one inflated by solvent molecules (Fig.~\ref{fig:MDmodels}(b) in SM~\cite{SMnote}).
Our result, instead, makes clear that the expansion is not a true thermodynamic transition but rather the result of the {\it continuous} modification of the coexistence line upon varying $\lambda$, with the system always remaining in a biphasic region.

\begin{figure}
\includegraphics[width=0.46\textwidth]{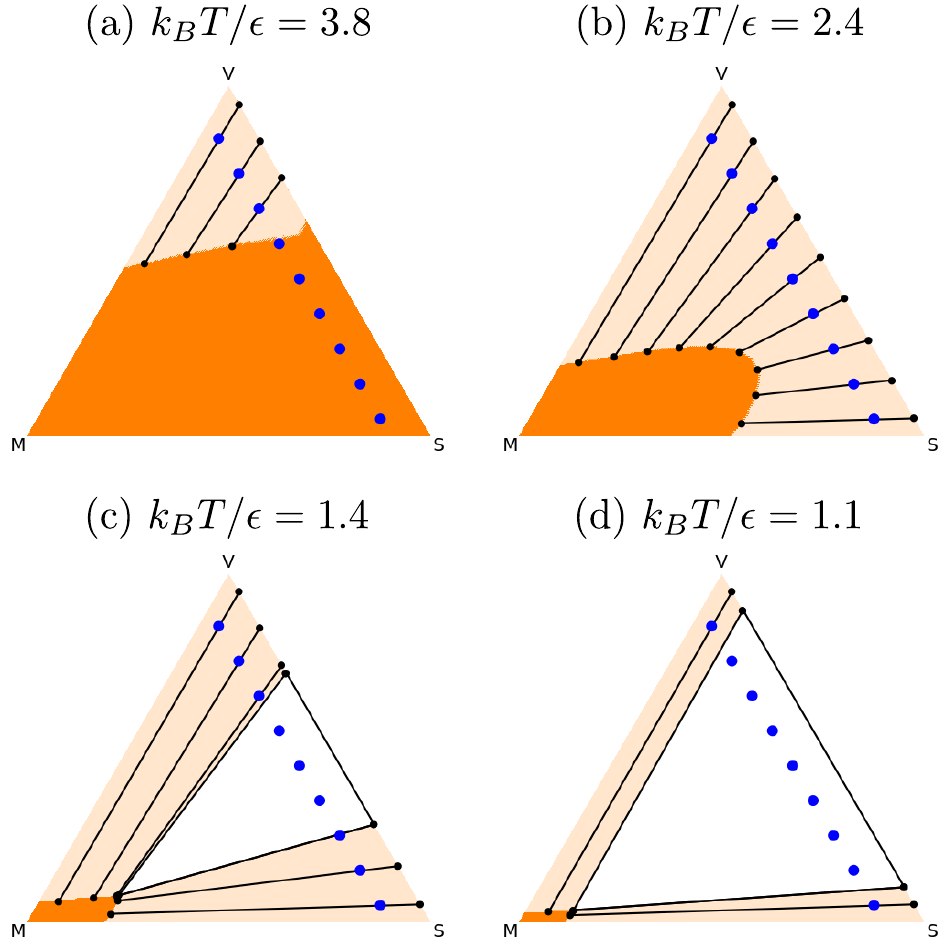}
\caption{
Phase behavior of a single chain in a one-solvent bath of fixed quality and upon varying temperature (Eq.~\eqref{eq:FreeEneExplicitSolvent} for $d=3$, $\epsilon_{mm}=\epsilon_{ss}=-\epsilon<0$, $\epsilon_{ms}=-0.7\epsilon$, $\epsilon_{cc}=\epsilon_{mc}=\epsilon_{sc}=0$ and selected values of $k_BT/\epsilon$). 
Symbols, color code and notation are as in Fig.~\ref{fig:MeanFieldResultsT2.5}.
At high temperatures (panels (a) and (b), system's behavior is as seen in the two-phase situation.
For $k_BT / \epsilon <1.5$ (panels (c) and (d)), triphasic stability (white triangular region, identified by the convex hull procedure) becomes possible: if the system is prepared inside this region, it separates into $3$ coexisting phases whose compositions (obtained by solving numerically Eqs.~\eqref{eq:CoexistenceEqsTriPhase-a}-\eqref{eq:CoexistenceEqsTriPhase-d}) lie at the corners of the white region.
}
\label{fig:MeanFieldResultsLambda0.7}
\end{figure}

Polymer collapse can be also studied by fixing the $ms$ interaction while decreasing the temperature.
Here, this translates in fixing $\lambda$ (we choose $\lambda=0.7$) and decreasing temperature $T$ (Fig.~\ref{fig:MeanFieldResultsLambda0.7}).
At high $T$ (panel (a)), the system separates in two phases only for very low $\phi_s$.
Then, as temperature drops (panel (b)), more and more points are incorporated in the two-phase region until, for $k_BT / \epsilon <1.5$ (see Sec.~\ref{sec:BinaryMixtures} in SM~\cite{SMnote}), a triphasic domain appears and extends progressively.
The latter corresponds to the white triangle in panels (c) and (d), whose corners identifies the stable compositions into which the system separates; one phase is a polymer globule that becomes more and more compact with decreasing temperature, while the other two are a solvent-poor and a solvent-rich phase.
Three-phase coexistence was {\it not} studied in past numerical simulations~\cite{Huang2021,Garg2023} and our analysis indicates that it ought to be seen at sufficiently low temperatures.

Notably, the picture and numerical methodologies described so far can be extended to systems of multiple chains with a finite mean contour length ($\ell<\infty$). 
Here one has to solve a larger number of equations characterizing equilibrium (see Sec.~\ref{sec:PhaseStability-MultiChains} in SM~\cite{SMnote}) in order to determine the coexistence lines, but otherwise there are no significant complications with respect to the single-chain case.
Also for a multi-chain system, our mean-field theory predicts re-entrant phase behavior for increasing $\lambda$ (Fig.~\ref{fig:MeanFieldResultsMultiChainl10} in SM~\cite{SMnote}) that differs from the single-chain case since the polymer-poor phase is now characterized by a small, yet non-zero, value of the monomer density.
In general, this is also in agreement with numerical simulations of polymer solutions~\cite{Huang2021}. 
Finally, as in the single-chain case, the temperature dependence is also of particular interest and was not analyzed numerically before.
Our theory predicts again that upon cooling there is a large region of the parameter space guaranteeing triphasic stability (white regions in Fig.~\ref{fig:MeanFieldResultsMultiChainl10Lambda0.7} in SM~\cite{SMnote}, especially panel (e)).
Notice that, being the multi-chain system polydisperse within the context of our model, we cannot rule out the possibility of polymer fractionation~\cite{Flory1953Principles}, namely that the mean chain length $\ell$ differs according to the phase.
For simplicity however, we do not consider this phenomenology here and postpone the discussion of the effects of polydispersity in future work. 

\subsection{
Polymer co-nonsolvency 
}\label{sec:Cononsolvency}
As a second relevant application to single-chain systems, we consider the phenomenon known as co-nonsolvency.
Here (see Fig.~\ref{fig:CoNonSolvency-Kremer} in SM~\cite{SMnote}), a polymer chain in a mixture formed by a solvent and a cosolvent (each of which, individually, would maintain the polymer in a swollen, coil-like conformation) undergoes, quite surprisingly, a re-entrant coil-globule-coil behavior upon systematically changing the concentration of the cosolvent~\cite{KremerNatCom2014,Freed2015,Sommer2018,Sommer2022cononsolvency,Zhang2024,Meng2024}. 

\begin{figure}
\includegraphics[width=0.46\textwidth]{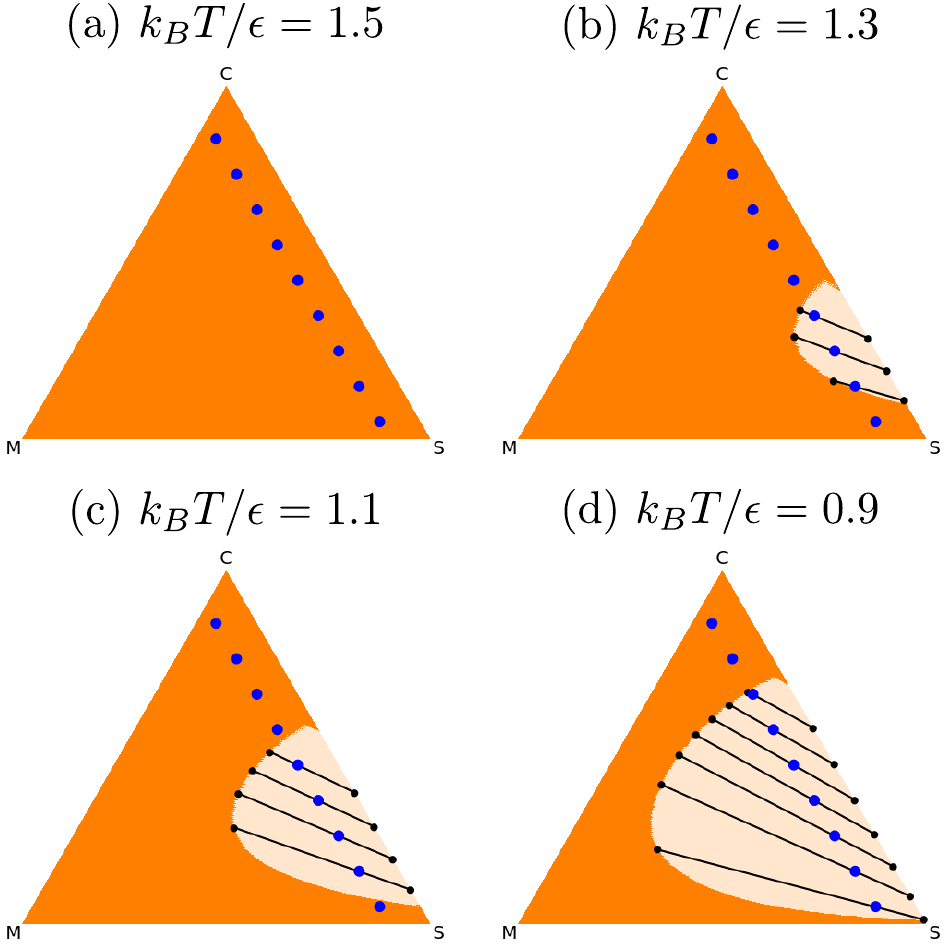}
\caption{
Phase behavior of co-nonsolvency
(Eq.~\eqref{eq:FreeEneExplicitSolvent} for $d=3$ and for all pair interactions set to $0$ except $\epsilon_{mc} = -\epsilon$ with $\epsilon > 0$, so mimicking the conditions of~\cite{KremerNatCom2014}).
Symbols, color code and notation are as in Fig.~\ref{fig:MeanFieldResultsT2.5}.
As $k_B T/\epsilon$ is lowered, a biphasic region appears and the chain collapses to a globule for a range of values of the cosolvent density $\phi_c$.
Reading the diagrams at fixed $k_B T/\epsilon$ and fixed $\phi_m$, the coil-globule-coil polymer behavior upon increasing $\phi_c$ as reported in~\cite{KremerNatCom2014} becomes apparent.
}
\label{fig:CononsolvencySingleChain}
\end{figure}

To capture the physics behind co-nonsolvency, it is important to realize that 
solvent and cosolvent are, in general, not perfectly equivalent and that, under suitable conditions, the second can constitute a better solvent for the polymer than the first.
In order to account for this mechanism, we follow the numerical work~\cite{KremerNatCom2014} and set in the free energy~\eqref{eq:FreeEneExplicitSolvent} all pair interactions $=0$ except for $\epsilon_{mc} = -\epsilon$ with $\epsilon > 0$.
Fig.~\ref{fig:CononsolvencySingleChain} illustrates the corresponding phase diagram of the system for different values of $k_B T/\epsilon$.
Below a certain critical value of $k_B T/\epsilon$ a biphasic region appears and the chain collapses to a globule.
When this happens, for a fixed value of $k_B T/\epsilon$ and for increasing values of $\phi_c$ the coil-globule-coil transition reported in~\cite{KremerNatCom2014} can be observed.
This is also in accordance with other works (see for instance~\cite{Freed2015,Zhang2024,Meng2024}) where the authors argued that the co-nonsolvency effect can in fact be explained in terms of a Flory-Huggins-like theory, challenging what appear to be other alternative views on the topic~\cite{KremerNatCom2014,Sommer2018,Sommer2022cononsolvency}.

\subsection{
Polymer-assisted condensation 
}\label{sec:PAC}
Finally, we consider 
the recent numerical study on ``polymer-assisted condensation''~\cite{SchiesselPAC2022} at the basis of the formation of biomolecular condensates within cell nuclei.
Here, phase separation of a two-component liquid mixture, otherwise in a single stable phase, is catalyzed by the presence of a single polymer chain that displays preferential attachment to one of the two components
(see Fig.~\ref{fig:PAC-Sommer+Schiessel} in SM~\cite{SMnote} for an illustration of the phenomenon). 

\begin{figure}
\includegraphics[width=0.46\textwidth]{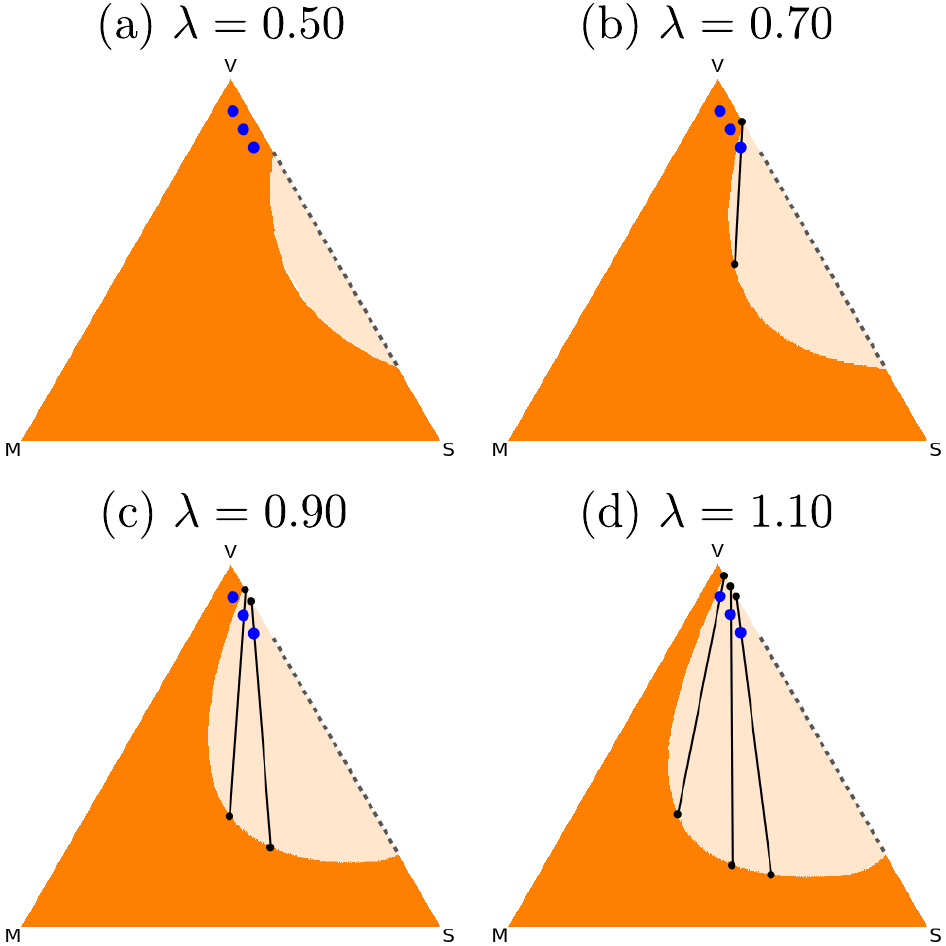}
\caption{
Phase behavior of polymer-assisted condensation (Eq.~\eqref{eq:FreeEneExplicitSolvent} in the main text for $d=3$, $\epsilon_{mm}=0$, $\epsilon_{ss}=-\epsilon<0$, $\epsilon_{ms}=-\lambda\epsilon$ ($\lambda>0$), $\epsilon_{cc}=\epsilon_{mc}=\epsilon_{sc}=0$ and $k_BT/\epsilon=1.3$). 
The gray dashed line denotes the {\it miscibility gap}~\cite{Rubinstein2003} of the binary $SV$ mixture: its extremities correspond to the binodal concentrations. 
The three large blue dots denote corresponding mean compositions of the system with the same monomer density $\phi_m = 0.04$ and as many values of $\phi_s$ outside the miscibility gap.
}
\label{fig:MeanFieldResultsPAC}
\end{figure}

Within our theoretical approach, the conditions of Ref.~\cite{SchiesselPAC2022} can be easily reproduced by
modeling, similar to Sec.~\ref{sec:Results-SingleChain}, the original two-component mixture as made of solvent molecules and vacancies.
Again then, in the free energy~\eqref{eq:FreeEneExplicitSolvent}
we fix the parameters $\epsilon_{mc} = \epsilon_{sc} = \epsilon_{cc} = 0$, while we assume 
purely steric $mm$ interactions ($\epsilon_{mm}=0$) and $ss$ and $ms$ attractions ($\epsilon_{ss}=-\epsilon<0$ and tuneable $\epsilon_{ms}=-\lambda\epsilon$ {\it via} $\lambda>0$).
By setting $k_BT / \epsilon =1.3$ ($<$ than the critical value $=1.5$ for phase separation in the solvent-vacancy binary system, see Sec.~\ref{sec:BinaryMixtures} in SM~\cite{SMnote}), we select 3 representative coordinates $(\phi_m, \phi_s, \phi_v)$ with solvent densities outside the so called {\it miscibility gap}~\cite{Rubinstein2003} of the binary solvent-vacancy set-up and study their stability by varying $\lambda$. 
Remarkably (Fig.~\ref{fig:MeanFieldResultsPAC}), our theory reproduces the simulation results of Ref.~\cite{SchiesselPAC2022}: in particular, for low $\lambda$ (panel (a)) all three points are stable, while as $\lambda$ increases (panels (b)-(d)) the early stable compositions phase-separate, beginning with those with the highest $\phi_s$.

\section{Discussion and conclusions}\label{sec:DiscConcl}
In this paper, we have introduced a new $O(n \to 0)$-vector spin model leading to the {\it exact} grand canonical partition function of lattice polymers with explicit solvent
and cosolvent 
molecules (Eq.~\eqref{eq:GranCanonicalZ}), and mapped it to a field-theoretic form that is amenable to a saddle-point approximation.
The Legendre transform of the resulting expression gives the mean-field free energy of the system (Eq.~\eqref{eq:FreeEneExplicitSolvent}) that generalizes earlier work~\cite{Doniach1996,Marcato2023} and, notably, is of the Flory-Huggins form~\cite{Flory1942,Huggins1942} for a {\it ternary} mixture of polymer, solvent
and cosolvent. 

A systematic stability analysis of the equilibrium relations for two (Eqs.~\eqref{eq:CoexistenceEqs1}-\eqref{eq:CoexistenceEqs3} and Eqs.~\eqref{eq:CoexistenceEqsMultiChain-a}-\eqref{eq:CoexistenceEqsMultiChain-d} in SM~\cite{SMnote}) and three phases (Eqs.~\eqref{eq:CoexistenceEqsTriPhase-a}-\eqref{eq:CoexistenceEqsTriPhase-d} and Eqs.~\eqref{eq:CoexistenceEqs1TriPhaseMC}-\eqref{eq:CoexistenceEqs4TriPhaseMC} in SM~\cite{SMnote}) coupled with the convex hull method~\cite{Wolff2011,Mao2019} reproduces transparently recent results from extensive numerical simulations for single- and multi-chain systems in explicit solvent~\cite{Huang2021,SchiesselPAC2022,Garg2023}.
In particular, we provide a {\it unified} explanation for
three, 
seemingly unrelated, observations:
(i)
a re-entrant polymer coil-globule transition (Fig.~\ref{fig:MeanFieldResultsT2.5}),
(ii)
polymer co-nonsolvency (Fig.~\ref{fig:CononsolvencySingleChain}), 
(iii)
polymer-assisted condensation of the solvent (Fig.~\ref{fig:MeanFieldResultsPAC}).

It is particularly worth emphasizing how in both (i) and (iii) cases the transition is triggered by the increasing $\epsilon_{ms}$-attraction between the polymer and the solvent, and yet the role of the third component (the vacancies in this case) remains essential.
As a matter of fact, neither of these two phenomena would occur in a pure polymer/solvent binary mixture.
More generally, our approach underscores the delicate ``solvent/cosolvent'' interplay and highlights the important regions in the temperature/density plane, thus providing a powerful guideline to specific numerical simulations.
In particular, we have unveiled whole new regions of low-temperature three-phase stability (Fig.~\ref{fig:MeanFieldResultsLambda0.7} and Fig.~\ref{fig:MeanFieldResultsMultiChainl10Lambda0.7} in SM~\cite{SMnote}) that were never studied before, to the best of our knowledge.

To conclude, we highlight some ideas for future work.
First, the main tools introduced here can be readily generalized to mixtures with more species, to polymers with intrinsic bending stiffness~\cite{Marcato2023} or with complex architectures~\cite{LubenskyPRA1979}.
Another potential avenue for this method concerns the so-called {\it crowding effect} or {\it depletion interaction}.
As the ratio between the size of the solvent and the size of the polymer decreases, a collapse of the polymer is expected below a certain asymmetry, purely for entropic reasons~\cite{Ye1996}.
Although a rigorous field-theoretic treatment of this problem along the lines of the present work appears far from trivial, it may still be possible to account for this effect even within a lattice model, and we plan to address this point in a future dedicated study. 
A final perspective is related to the extension of the present theory beyond mean-field~\cite{deGennes1972,GabayGarel-RG-1978}, particularly with the inclusion of finite-size effects~\cite{LifshitzRMP1978} that become important in the vicinity of continuous transitions, and the study of the kinetics of phase separation~\cite{Mao2019}.

{\it Acknowledgements} --
A.G. acknowledges the MIUR PRIN-COFIN2022 grant 2022JWAF7YMIUR. 
A.G. and A.R. acknowledge networking support by the COST Action CA17139 (EUTOPIA).
D.M. and A.R. acknowledge H. Schiessel for fruitful discussions, in particular for bringing to our attention the recent work on polymer-assisted condensation.





\bibliography{../biblio}

\begin{thebibliography}{65}
\bibitem [{Eas()}]{EasyOnLimit}%
  \BibitemOpen
  \href@noop {} {}\bibinfo {note} {Specifically, by this ansatz every
  dependence on $n$ disappears, therefore the limit $n \to 0$ becomes
  trivial.}\BibitemShut {Stop}%
\bibitem [{NoP()}]{NoPolymerChemPotentialNote}%
  \BibitemOpen
  \href@noop {} {}\bibinfo {note} {Notice that where one phase is characterized
  by the absence of one species (as it is the case of monomers here), the
  expression for the equilibrium of the related chemical potential is
  absent~\cite{deGennesBook}. Instead, the balance of chemical potentials
  between different monomer phases needs to be enforced for multi-chain systems
  (see Sec.~\ref{sec:PhaseStability-MultiChains} in
  SM~\cite{SMnote}).}\BibitemShut {Stop}%
\end{thebibliography}

\clearpage

\widetext
\clearpage
\begin{center}
\textbf{\Large Supplemental Material \\ \vspace*{1.5mm} Theory of polymers in binary solvent solutions:\\ mean-field free energy and phase behavior} \\
\vspace*{5mm}
Davide Marcato, Achille Giacometti, Amos Maritan, Angelo Rosa
\vspace*{10mm}
\end{center}

\setcounter{equation}{0}
\setcounter{figure}{0}
\setcounter{table}{0}
\setcounter{page}{1}
\setcounter{section}{0}
\setcounter{page}{1}
\makeatletter
\renewcommand{\theequation}{S\arabic{equation}}
\renewcommand{\thefigure}{S\arabic{figure}}
\renewcommand{\thetable}{S\arabic{table}}
\renewcommand{\thesection}{S\arabic{section}}
\renewcommand{\thepage}{S\arabic{page}}

\tableofcontents

\clearpage

\section{Binary mixtures as special cases}\label{sec:BinaryMixtures}
It is instructive to specialize the free energy $\beta f$ (Eq.~\eqref{eq:FreeEneExplicitSolvent} in the main text) to {\it binary} mixtures, namely when one of the three considered species (monomers ($m$), solvent molecules ($s$) or cosolvent molecules ($c$)) is absent.
In particular, after subtracting the contribution of the free energy of the corresponding phase-separated system, one gets the functional form of the free energy of {\it mixing} and the related {\it Flory parameter}, $\chi$, recapitulating the interaction between the two species (for reference, see~\cite{deGennesBook,Rubinstein2003}).
Essentially, two situations are possible:
\begin{enumerate}
\item $\phi_c=1-\phi_m-\phi_s=0$:
$\beta f(\phi_m, \ell) = d \beta \left(\epsilon_{mm}-2\epsilon_{ms}+\epsilon_{ss}\right) \, \phi_m^2  + \frac{\phi_m}{\ell} \ln(\phi_m) + (1-\phi_m) \ln(1-\phi_m) + \phi_m\left[\ln\!\left( \frac{(1-2/\ell)^{1-2/\ell} \, (2/{\ell}^2)^{1/\ell}}{(2d \, e^{\beta\epsilon_{mm}-1} (1-1/\ell))^{1-1/\ell}} \right)+2d\beta\left(\epsilon_{ms}-\epsilon_{ss}\right)\right] + d\beta\epsilon_{ss}$ corresponds to the free energy of a mixture of polymer chains of mean contour length $\ell$ and solvent molecules.
Such a mixture is represented as a point on the $MS$ side of the Gibbs triangle (Fig.~\ref{fig:GibbsTriangle} in the main text).

By the free energy of mixing -- $\beta \Delta f(\phi_m) \equiv \beta f(\phi_m) - \phi_m\beta f(\phi_m=1) - (1-\phi_m)\beta f(\phi_m=0)$ -- the Flory parameter for the polymer/solvent mixture is given by $\chi = d\beta(2\epsilon_{ms}-\epsilon_{mm}-\epsilon_{ss})$.
By standard methods, it is easy to derive the critical temperatures for phase separation for the two following cases:
(i) for $\ell=1$, $T^{*}=\frac{d(2\epsilon_{ms}-\epsilon_{mm}-\epsilon_{ss})}{2k_B}$;
(ii) for $\ell \to \infty$, $T^{*}=\frac{2d(2\epsilon_{ms}-\epsilon_{mm}-\epsilon_{ss})}{k_B}$.

With reference to the case of Fig.~\ref{fig:MeanFieldResultsT2.5} in the main text ($\ell\to\infty$, $d=3$, $\epsilon_{mm}=\epsilon_{ss}=-\epsilon<0$ and $\epsilon_{ms}=-\lambda\epsilon$ with $\lambda>0$), the critical temperature $\frac{k_BT^{*}}{\epsilon} = 12(1-\lambda)$.
The system may then phase-separate only for $\lambda < 1$ while for $\lambda>1$ the entire $MS$ side is in the stable region regardless of the value of the temperature $T$, as confirmed by the figure.

Obviously, the situation in which $\phi_s = 0$ (and thus $\phi_c = 1-\phi_m$) is specular to the one just discussed, so one gets $\beta f(\phi_m, \ell) = d \beta \left(\epsilon_{mm}-2\epsilon_{mc}+\epsilon_{cc}\right) \, \phi_m^2  + \frac{\phi_m}{\ell} \ln(\phi_m) + (1-\phi_m) \ln(1-\phi_m) + \phi_m\left[\ln\!\left( \frac{(1-2/\ell)^{1-2/\ell} \, (2/{\ell}^2)^{1/\ell}}{(2d \, e^{\beta\epsilon_{mm}-1} (1-1/\ell))^{1-1/\ell}} \right)+2d\beta\left(\epsilon_{mc}-\epsilon_{cc}\right)\right] + d\beta\epsilon_{cc}$, which is identical to the previous expression except for the fact that the solvent $s$ has been replaced by the cosolvent $c$.
\item $\phi_m=0$:
$\beta f = \beta f(\phi_s) = d \beta (\epsilon_{ss}-2\epsilon_{sc}+\epsilon_{cc})\, \phi_s^2 + (1 - \phi_s) \ln(1 - \phi_s) + \phi_s \ln(\phi_s) + 2d\beta(\epsilon_{sc}-\epsilon_{cc})\phi_s + d\beta\epsilon_{cc}$ corresponds to the free energy of a binary mixture of solvent and cosolvent molecules.
Such a mixture is represented as a point on the $CS$ side of the Gibbs triangle (Fig.~\ref{fig:GibbsTriangle} in the main text).

By the free energy of mixing -- $\beta \Delta f(\phi_s) \equiv \beta f(\phi_s) - \phi_s\beta f(\phi_s=1) - (1-\phi_s)\beta f(\phi_s=0)$ -- the Flory parameter for the solvent molecules is $\chi = d\beta(2\epsilon_{sc} - \epsilon_{ss}-\epsilon_{cc})$.
Accordingly, phase separation takes place at the {\it critical} Flory parameter $\chi^{*}=2$, corresponding to the critical temperature $T^{*}=\frac{d(2\epsilon_{sc}-\epsilon_{ss}-\epsilon_{cc})}{2k_B}$ (obviously, physical values of $T^{*}$ are possible only for $\chi > 0$, namely when the effective interaction among solvent molecules is attractive).

With reference to the case of Fig.~\ref{fig:MeanFieldResultsLambda0.7} in the main text ($d=3$ and $\epsilon_{ss}=-\epsilon<0$), the critical temperature $\frac{k_BT^{*}}{\epsilon} = \frac32$: accordingly, for $T<T^{*}$ some points on the $SC$ side become non-stable and a triphasic region appears.

\end{enumerate}
%

\section{Phase-stability: multi-chain systems}\label{sec:PhaseStability-MultiChains}
In this Section, we describe the conditions for two-phase (Sec.~\ref{sec:TwoPhase-MultiChain}) and three-phase (Sec.~\ref{sec:3Phase-MultiChain}) stability for a polymer solution in explicit solvent, assuming chains of {\it finite} mean contour length ({\it i.e.}, $\ell < \infty$ in Eq.~\eqref{eq:FreeEneExplicitSolvent} in the main text).
Now, contrary to the single-chain case, in the derivation of the equations for the equilibrium densities (namely, the equivalent of Eqs.~\eqref{eq:CoexistenceEqs1}-\eqref{eq:CoexistenceEqs3} and Eqs.~\eqref{eq:CoexistenceEqsTriPhase-a}-\eqref{eq:CoexistenceEqsTriPhase-d} in the main text) we take $\phi_m^I > 0$.
As a consequence, in addition to the balance of the chemical potential of the solvent (see Eq.~\eqref{eq:DefineChemicalPot} in the main text), we have here one additional equilibrium condition to impose coming from the balance of the monomer chemical potential,
\begin{equation}\label{eq:DefineChemicalPot-Monomers}
\mu_m(\phi_m, \phi_s) \equiv \frac{\partial f}{\partial \phi_m} \, .
\end{equation}
%

\subsection{Two-phase stability}\label{sec:TwoPhase-MultiChain}
By adopting the same notation of the main text and with the additional constraint from Eq.~\eqref{eq:DefineChemicalPot-Monomers}, the new set of equilibrium equations becomes:
\begin{eqnarray}
\Pi(\phi^{I}_m, \phi^{I}_s) & = & \Pi(\phi^{II}_m, \phi^{II}_s) \, , \label{eq:CoexistenceEqsMultiChain-a} \\
\mu_s(\phi^{I}_m, \phi_s^I) & = & \mu_s(\phi^{II}_m, \phi_s^{II}) \, ,
\label{eq:CoexistenceEqsMultiChain-c} \\
\mu_m(\phi_m^I, \phi_s^I) & = & \mu_m(\phi_m^{II}, \phi_s^{II}) \, , \label{eq:CoexistenceEqsMultiChain-b} \\
\frac{\phi_m - \phi^{I}_m}{\phi^{II}_m - \phi^{I}_m} & = & \frac{\phi_s - \phi^{I}_s}{\phi^{II}_s - \phi^{I}_s} \, . \label{eq:CoexistenceEqsMultiChain-d}
\end{eqnarray}
Notice in particular, and in comparison to Eqs.~\eqref{eq:CoexistenceEqs1}-\eqref{eq:CoexistenceEqs3} in the main text, the ``new'' Eq.~\eqref{eq:CoexistenceEqsMultiChain-b} as the consequence of having $\phi_m^I>0$.
Moreover, and as already noticed for $\ell \to \infty$, also in this more general case the solution to Eqs.~\eqref{eq:CoexistenceEqsMultiChain-a}-\eqref{eq:CoexistenceEqsMultiChain-d} depends explicitly on the preparation conditions of the system, $\phi_m$ and $\phi_s$.

Two-phase stability for multi-chain systems is illustrated in Fig.~\ref{fig:MeanFieldResultsMultiChainl10} for the same set of parameters considered in Fig.~\ref{fig:MeanFieldResultsT2.5} in the main text and chain mean length $\ell=10$.
The major difference with respect to the single-chain case is in the appearance of a dilute, yet strictly $>0$, polymer phase that, intuitively, is expected to become more pronounced for lower values of $\ell$.
Finally, as it can be noticed in Fig.~\ref{fig:MeanFieldResultsMultiChainl10}, a re-entrance condition does still exist for $\lambda > 1.0$, in agreement with the results of molecular dynamics computer simulations by Huang and Cheng~\cite{Huang2021}.

\subsection{Three-phase stability}\label{sec:3Phase-MultiChain}
The last case that needs to be addressed is that of three-phase coexistence in multi-chain systems.
Again, the main difference with respect to the single-chain counterpart is that we can no longer assume that $\phi_m = 0$ in any of the phases in which the system separates.
Therefore, in total we need now to compute $6$ equilibrium densities.
By implementing the same procedure of the minimization of the free energy (Eq.~\eqref{eq:FreeEneExplicitSolvent} in the main text) with the proper constraints leads to the following set of equations:
\begin{eqnarray}
\Pi(\phi_m^I, \phi_s^I) & = & \Pi(\phi_m^{II}, \phi_s^{II}) \, , \label{eq:CoexistenceEqs1TriPhaseMC} \\
\Pi(\phi_m^I, \phi_s^I) & = & \Pi(\phi_m^{III}, \phi_s^{III}) \, , \\
\mu_s(\phi_m^I, \phi_s^I) & = & \mu_s(\phi_m^{II}, \phi_s^{II}) \, , \label{eq:CoexistenceEqs5TriPhaseMC} \\
\mu_s(\phi_m^I, \phi_s^I) & = & \mu_s(\phi_m^{III}, \phi_s^{III}) \, ,
\label{eq:CoexistenceEqs6TriPhaseMC} \\
\mu_m(\phi_m^I, \phi_s^I) & = & \mu_m(\phi_m^{II}, \phi_s^{II}) \, , \label{eq:CoexistenceEqs3TriPhaseMC} \\
\mu_m(\phi_m^I, \phi_s^I) & = & \mu_m(\phi_m^{III}, \phi_s^{III}) \, . 
\label{eq:CoexistenceEqs4TriPhaseMC}
\end{eqnarray}
Once again, we point out that Eqs.~\eqref{eq:CoexistenceEqs1TriPhaseMC}-\eqref{eq:CoexistenceEqs4TriPhaseMC} (to be compared to Eqs.~\eqref{eq:CoexistenceEqsTriPhase-a}-\eqref{eq:CoexistenceEqsTriPhase-d} in the main text) imply the equality of the osmotic pressure and the chemical potentials of both species (monomers and solvent molecules) in all 3 phases.

Three-phase stability for multi-chain systems is illustrated in Fig.~\ref{fig:MeanFieldResultsMultiChainl10Lambda0.7} for the same set of parameters considered in Fig.~\ref{fig:MeanFieldResultsLambda0.7} in the main text and chain mean length $\ell=10$.
Qualitatively the behavior appears similar to the single-chain case, the major difference is (as in two-phase behavior) the appearance of a dilute, strictly $>0$, polymer phase.

\clearpage
\section*{Supplemental figures}
\clearpage

\begin{figure*}
\includegraphics[width=0.95\textwidth]{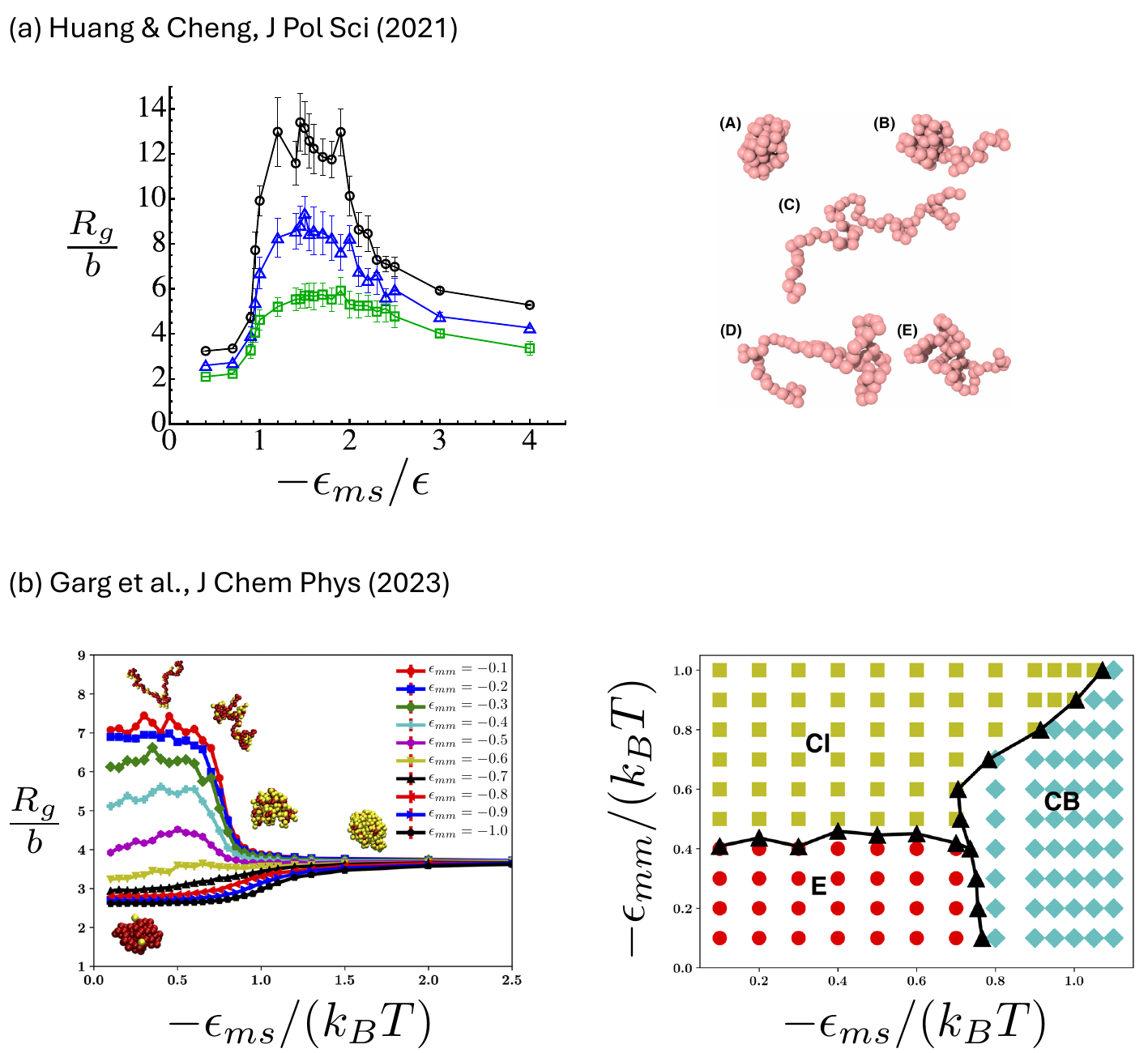}
\caption{
Phase-behavior of a single polymer chain in a one-solvent bath, insights from Molecular Dynamics computer simulations.
Polymers are modeled as linear chains of beads (or monomers) and solvent molecules as single particles; monomers and solvent particles have the same linear size $=b$, and monomer-monomer ($mm$), solvent-solvent ($ss$) and monomer-solvent ($ms$) interactions are of the Lennard-Jones type.
(a)
The first set-up is from Ref.~\cite{Huang2021}, with a single chain immersed in a bath of solvent molecules at density $\rho_s = 0.64 / b^3$.
$mm$ and $ss$ pair interactions are attractive, with equal fixed strength defining our energy scale ($\epsilon_{mm}=\epsilon_{ss}=-\epsilon<0$); the $ms$ interaction is attractive and of variable strength.
The polymer mean gyration radius in bond units ($R_g / b$, l.h.s.) as a function of the $ms$ strength ($-\epsilon_{ms}/\epsilon$) demonstrates that the polymer undergoes a {\it compact-swollen-compact} re-entrant behavior upon {\it increasing} the polymer affinity with the solvent.
The re-entrant globule phase is characterized by the presence of solvent molecules inside the globule, which makes it more swollen.
On the r.h.s., a few representative chain conformations from low to high $ms$ affinity (panels (A) to (E)).
Notice that in the same work, the authors discussed also multi-chain systems.
{\bf Reprinted and adapted from [Y. Huang, S. Cheng, Journal of Polymer Science 59, 2819 (2021); Ref.~\cite{Huang2021}], with the permission of John Wiley and Sons.}
(b)
The second set-up is from Ref.~\cite{Garg2023}, with a single chain immersed in a bath of solvent molecules at density $\rho_s = 0.047 / b^3$ ({\it i.e.}, very dilute conditions and, so, much smaller than in the previous case).
The $ss$ interaction strength ($\epsilon_{ss}=-\epsilon$) fixes the energy scale, and $mm$ ($\epsilon_{mm}$) and $ms$ ($\epsilon_{ms}$) are varied to characterize the mean chain gyration radius (l.h.s.) and the phase diagram (r.h.s.).
For relevant parameters here ($\epsilon_{mm}=\epsilon_{ss}$, bottom line in the l.h.s. panel), by increasing the $ms$ affinity the chain undergoes a transition from two distinct compact phases, from one (CI) stabilized by intra-polymer interactions to that (CB) stabilized by bridging (solvent-mediated) interactions.
{\bf Reprinted and adapted from [H. Garg et al., Journal of Chemical Physics 158, 114903 (2023); Ref.~\cite{Garg2023}], with the permission of AIP Publishing.}
}
\label{fig:MDmodels}
\end{figure*}

\begin{figure*}
$$
\begin{array}{ccc}
{\rm (a)} \, \lambda = 0.3 & {\rm (b)} \, \lambda = 0.5 & {\rm (c)} \, \lambda = 0.6 \\
\includegraphics[width=0.32\textwidth]{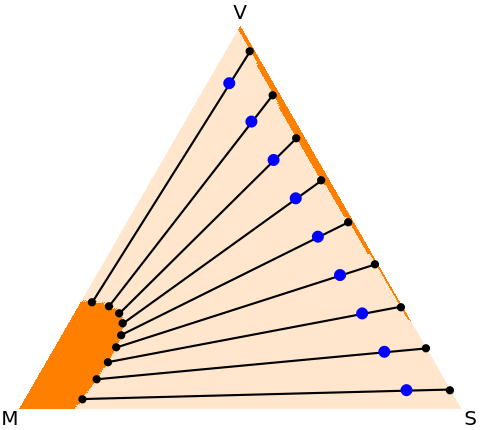} & \includegraphics[width=0.32\textwidth]{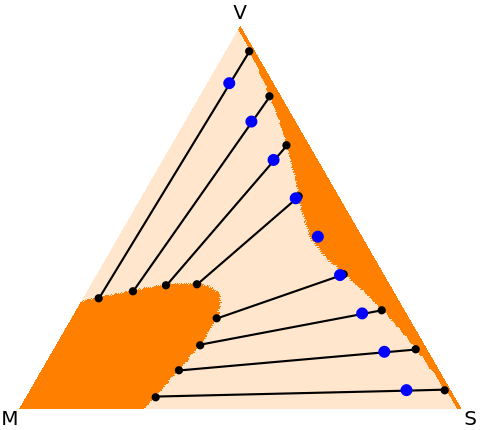} & \includegraphics[width=0.32\textwidth]{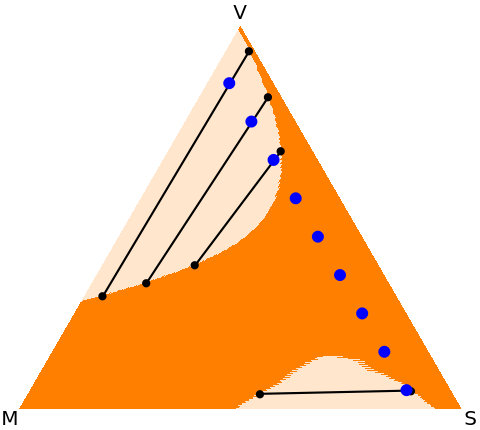} \\
& & \\
{\rm (d)} \, \lambda = 1.0 & {\rm (e)} \, \lambda = 1.3 & {\rm (f)} \, \lambda = 1.7 \\
\includegraphics[width=0.32\textwidth]{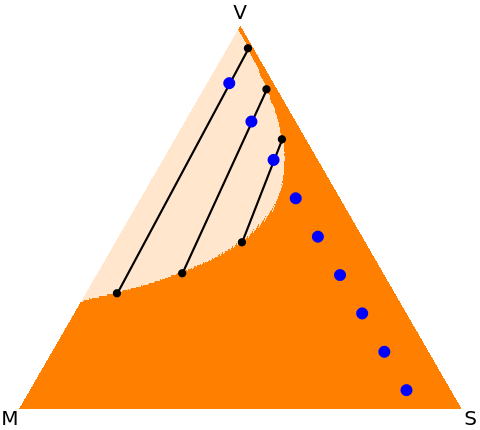} & \includegraphics[width=0.32\textwidth]{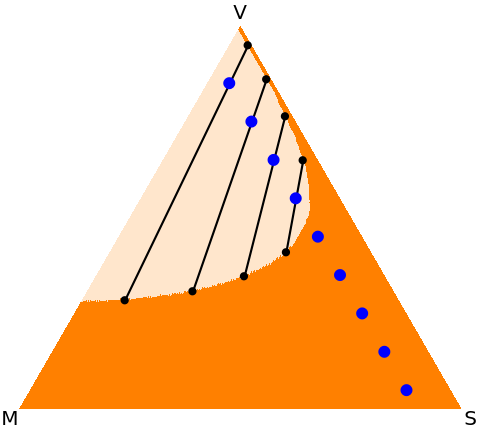} & \includegraphics[width=0.32\textwidth]{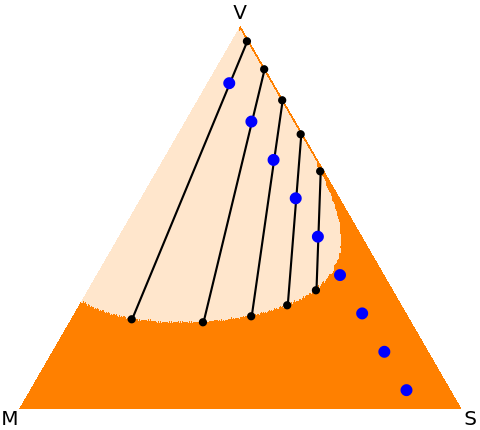}
\end{array}
$$
\caption{
Multi-chain phase behavior in a one-solvent bath of varying quality (Eq.~\eqref{eq:FreeEneExplicitSolvent} in the main text for chains of finite mean contour length $\ell=10$, and for $d$, $\epsilon_{mm}$, $\epsilon_{ss}$, $\epsilon_{cc}$, $\epsilon_{ms}$ $\epsilon_{mc}$, $\epsilon_{sc}$ and $T$ defined as in the caption of Fig.~\ref{fig:MeanFieldResultsT2.5} in the main text).
Symbols (in particular, the blue points corresponding to 9 chosen mean compositions of the system with the same $\phi_m=0.1$), color code and notation are as in Fig.~\ref{fig:MeanFieldResultsT2.5} in the main text.
The positions of the black dots are calculated by solving numerically Eqs.~\eqref{eq:CoexistenceEqsMultiChain-a}-\eqref{eq:CoexistenceEqsMultiChain-d}.
}
\label{fig:MeanFieldResultsMultiChainl10}
\end{figure*}

\begin{figure*}
$$
\begin{array}{ccc}
{\rm (a)} \, k_BT / \epsilon = 3.0 & {\rm (b)} \, k_BT / \epsilon = 2.3 & {\rm (c)} \, k_BT / \epsilon = 1.8 \\
\includegraphics[width=0.32\textwidth]{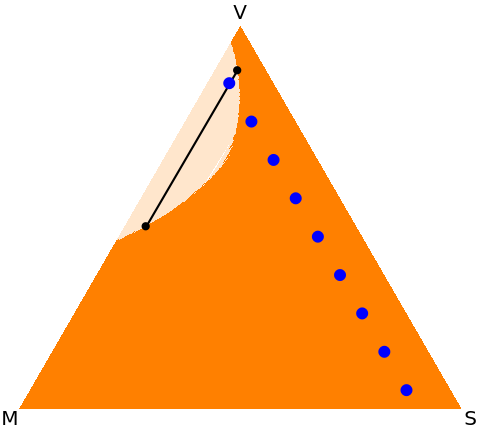} & \includegraphics[width=0.32\textwidth]{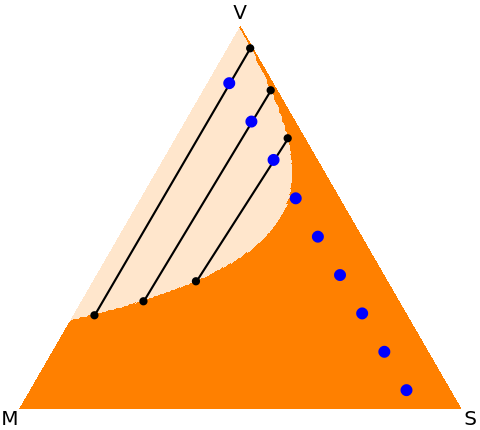} & \includegraphics[width=0.32\textwidth]{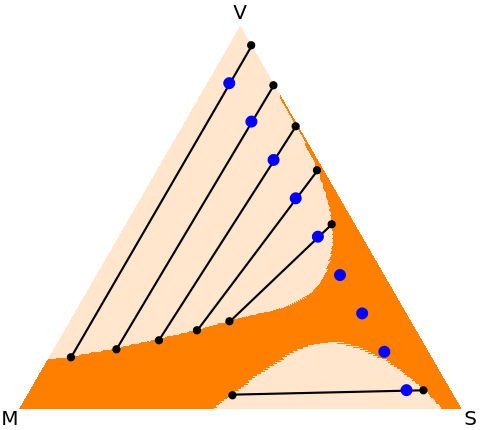} \\
& & \\
{\rm (d)} \, k_BT / \epsilon = 1.6 & {\rm (e)} \, k_BT / \epsilon = 1.4 & {\rm (f)} \, k_BT / \epsilon = 1.1 \\
\includegraphics[width=0.32\textwidth]{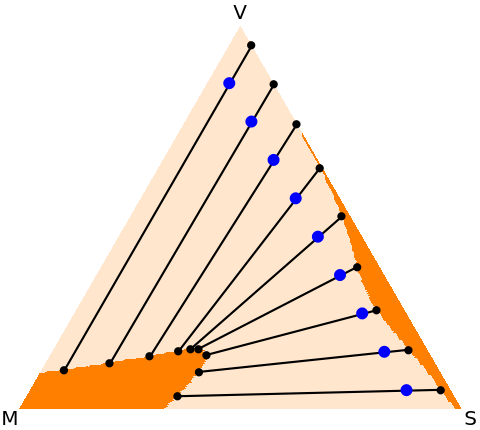} & \includegraphics[width=0.32\textwidth]{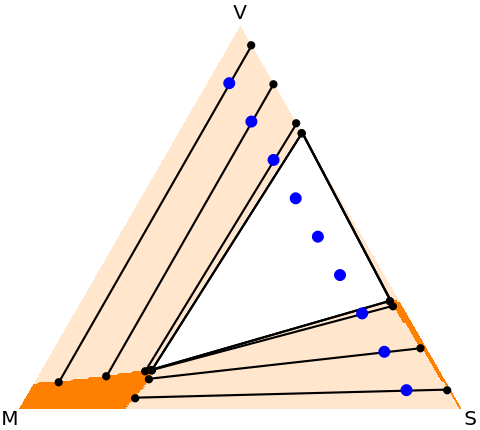} & \includegraphics[width=0.32\textwidth]{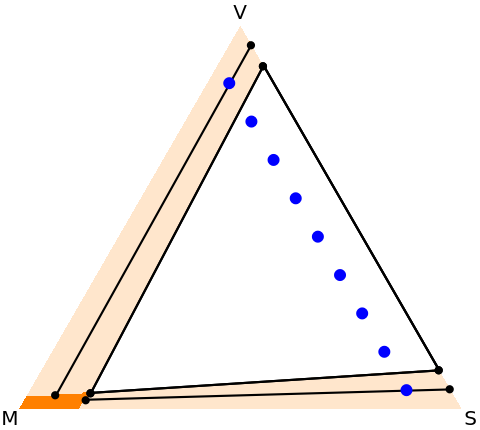}
\end{array}
$$
\caption{
Multi-chain phase behavior in a one-solvent bath of fixed quality and upon varying temperature (Eq.~\eqref{eq:FreeEneExplicitSolvent} in the main text for chains of finite mean contour length $\ell=10$, for $d$, $\epsilon_{mm}$, $\epsilon_{ss}$, $\epsilon_{cc}$, $\epsilon_{ms}$, $\epsilon_{mc}$, $\epsilon_{sc}$ defined as in the caption of Fig.~\ref{fig:MeanFieldResultsLambda0.7} in the main text and for selected values of $k_BT/\epsilon$).
At high temperatures (panels (a) to (c)) the behavior is similar to the single-chain situation, with larger portions of the Gibbs triangle interested by two-phase separation as temperature drops.
As temperature drops below the critical value (panels (d) to (e)), the triphasic region (white triangular region) becomes stable: the system at any mean composition inside this region separates into $3$ coexisting phases of compositions (obtained by solving numerically Eqs.~\eqref{eq:CoexistenceEqs1TriPhaseMC}-\eqref{eq:CoexistenceEqs6TriPhaseMC}) lying at the corners of the white triangle.
Symbols (in particular, the blue points corresponding to 9 chosen mean compositions of the system with the same $\phi_m=0.1$), color code and notation are as in Fig.~\ref{fig:MeanFieldResultsT2.5} in the main text.
}
\label{fig:MeanFieldResultsMultiChainl10Lambda0.7}
\end{figure*}

\begin{figure*}
\includegraphics[width=0.60\textwidth]{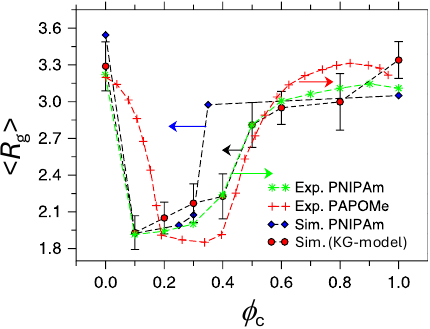}
\caption{
Phase-behavior of co-nonsolvency.
A polymer chain in a binary mixture of a solvent and a cosolvent (both good solvents for the polymer, but with the cosolvent a better solvent than the solvent) undergoes a coil-globule-coil phase behavior (namely, the mean gyration radius of the chain, $\langle R_g\rangle$, first decreases and then increases again) as a function of the density of the co-solvent, $\phi_c$.
This behavior is seen in experiments (PNIPAm and PAPOMe data) and well reproduced in conventional computer simulations, either full-atom (PNIPAm) or more coarse-grain (KG-model) ones.
{\bf Reprinted and adapted from [D. Mukherji {\it et al.}, Nature Communications 5, 4882 (2014); Ref.~\cite{KremerNatCom2014}] (the work is licensed under a Creative Commons Attribution-NonCommercial-NoDerivs 4.0 International License).}
}
\label{fig:CoNonSolvency-Kremer}
\end{figure*}

\begin{figure*}
\includegraphics[width=0.40\textwidth]{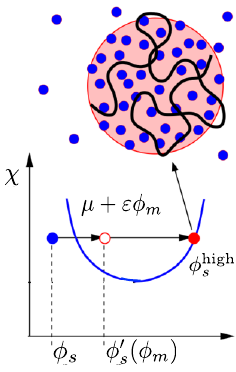}
\caption{
Phase-behavior of ``polymer-assisted condensation''.
In a binary (solvent/polymer) mixture, the solvent is set at concentration $\phi_s$ (blue dot) outside the region (delimited by the blue line) where, for the given interaction parameter ($\chi$), it would demix.
However, the presence of a favorable interaction with a polymer chain with monomer concentration $\phi_m$ shifts (locally) the concentration to a higher value ($\phi_s'(\phi_m)$, empty dot), enough to let formation of a high-concentration phase ($\phi_s^{\rm high}$, red dot) to happen.
{\bf Reprinted and adapted with permission from [J.-U. Sommer {\it et al.}, Macromolecules 55, 4841 (2022); Ref.~\cite{SchiesselPAC2022}]. Copyright 2022 American Chemical Society.}
}
\label{fig:PAC-Sommer+Schiessel}
\end{figure*}

\end{document}